\documentclass[traditabstract]{aa} 

\usepackage{graphicx}

\usepackage{lscape}
\usepackage{amssymb}
\usepackage{amsmath}

\usepackage{txfonts}
\usepackage{multirow}
\usepackage{booktabs}
\usepackage{subfigure}
\usepackage{placeins}
\usepackage{verbatim}
\usepackage{adjustbox}
\usepackage[toc,page]{appendix} 
\usepackage{tabularx}
\usepackage{lscape}
\usepackage{textcomp}
\usepackage{gensymb}
\usepackage{hyperref}
\usepackage{multicol}
\usepackage[symbol]{footmisc}
\usepackage{tikz}
\usepackage{caption}
\usepackage{longtable}
\usepackage{xtab}

\usepackage{natbib}

\usepackage{color}
\usepackage{xcolor}

\begin{document}

\title{Cosmology with supernova Encore in the strong lensing cluster MACS J0138$-$2155}
\subtitle{Spectroscopy with MUSE}
\authorrunning{G. Granata et al.}
\titlerunning{Lensing cluster of SN Encore with MUSE}

\author{G.~Granata\inst{\ref{unimi},\ref{unife},\ref{icg}}, 
        G.~B.~Caminha\inst{\ref{mpa},\ref{tum}},
        S.~Ertl\inst{\ref{mpa},\ref{tum}},
        C.~Grillo\inst{\ref{unimi},\ref{iasf}},
        S.~Schuldt\inst{\ref{unimi},\ref{iasf}},
        S.~H.~Suyu\inst{\ref{tum},\ref{mpa}},
        A.~Acebron\inst{\ref{ifca}},
        P.~Bergamini\inst{\ref{unimi},\ref{inafoas}},
        R.~Cañameras\inst{\ref{lam}},
        A.~M.~Koekemoer\inst{\ref{stsci}},
        P.~Rosati\inst{\ref{unife},\ref{inafoas}},
        \and
        S.~Taubenberger\inst{\ref{mpa},\ref{tum}}
}

\institute{
    Dipartimento di Fisica, Universit\`a  degli Studi di Milano, via Celoria 16, 20133 Milano, Italy \label{unimi}
    \and
    Dipartimento di Fisica e Scienze della Terra, Universit\`a degli Studi di Ferrara, via Saragat 1, 44122 Ferrara, Italy \label{unife}
    \and
    Institute of Cosmology and Gravitation, University of Portsmouth, Burnaby Rd, Portsmouth PO1 3FX, UK \label{icg}
    \\
    e-mail: \href{mailto:giovanni.granata@port.ac.uk}{\tt giovanni.granata@port.ac.uk}
    \and
    Max-Planck-Institut f{\"u}r Astrophysik, Karl-Schwarzschild Stra{\ss}e 1, 85748 Garching, Germany \label{mpa}
    \and
    Technical University of Munich, TUM School of Natural Sciences, Physics Department,  James-Franck-Stra{\ss}e 1, 85748 Garching, Germany \label{tum}
    \and 
    INAF -- IASF Milano, via Corti 12, 20133 Milano, Italy \label{iasf}
    \and
    Instituto de Física de Cantabria (CSIC-UC), Avda. Los Castros s/n, 39005 Santander, Spain \label{ifca}
    \and
    INAF – OAS, Osservatorio di Astrofisica e Scienza dello Spazio di Bologna, via Gobetti 93/3, 40129 Bologna, Italy \label{inafoas}
    \and
    Aix Marseille Univ, CNRS, CNES, LAM, Marseille, France \label{lam}
    \and
    Space Telescope Science Institute, 3700 San Martin Drive, Baltimore, MD 21218, USA \label{stsci}
}

\date{Received 17 Dec 2024; accepted 10 Mar 2025}

\abstract{We present a comprehensive spectroscopic analysis of MACS J0138$-$2155, at $z=0.336$.\ It is the first galaxy cluster known to host two strongly lensed supernovae (SNe), Requiem and Encore, and thus provides us with a chance to obtain a reliable $H_0$ measurement from the time delays between multiple images. We took advantage of new spectroscopic data from the Multi Unit Spectroscopic Explorer (MUSE) on the Very Large Telescope, complemented with archival imaging from the \textit{Hubble} Space Telescope. The MUSE data cover a central $1 \rm \, arcmin^2$ of the lensing cluster, for a total depth of 3.7 hours, including 2.9 hours recently obtained by our Target of Opportunity programme. Based on these observations, we release a new spectroscopic catalogue containing reliable redshifts for 107 objects. This includes 50 galaxy cluster members with secure redshift values in the range $0.324 < z < 0.349$, and 13 lensed multiple images from four background sources between $0.767\leq z \leq 3.420$, including four images of the host galaxy of the two SNe. We exploited the MUSE data to study the stellar kinematics of 14 bright cluster members and two background galaxies, obtaining reliable measurements of their line-of-sight velocity dispersion. Finally, we combined these results with measurements of the total magnitude of the cluster members in the \textit{Hubble} Space Telescope F160W band to calibrate the Faber-Jackson relation between luminosity and stellar velocity dispersion ($L \propto \sigma_v^{1/\alpha}$) for the early-type cluster member galaxies, measuring a slope $\alpha=0.25^{+0.05}_{-0.05}$  and a scatter of $25^{+6}_{-4} \, \rm km \, s^{-1}$ about the relation. We compared the calibrated relation with six other strong lensing clusters in the redshift range $0.31\leq z \leq 0.59$, finding a general agreement and no clear sign of evolution with redshift. A pure and complete sample of cluster member galaxies and a reliable characterisation of their total mass structure are key to building accurate total mass maps of the cluster and mitigating the impact of parametric degeneracies, which is necessary for inferring
the value of $H_0$ from the measured time delays between the lensed images of the two SNe.}


\maketitle
%
%________________________________________________________________
\section{Introduction}
\label{sec:intro}

Strong gravitational lensing (SL) by galaxy clusters has recently emerged as a unique bridge between astrophysics and cosmology. Galaxy clusters are remarkable as the most massive gravitationally bound structures in the Universe, with a mass budget dominated by dark matter (DM). Strong gravitational lensing is the most accurate probe of the total mass in the cores of massive clusters; when combined with tracers of the baryonic mass, it allows us to disentangle the mass distribution of the DM haloes \citep[e.g.][]{natarajan97,caminha19,granata22}. As such, SL models of galaxy clusters have been instrumental in unveiling and tackling some crucial open problems in state-of-the-art cosmology \citep[e.g.][]{grillo15,grillo18,meneghetti20}.

Galaxy clusters are privileged tracers of the formation and evolution of massive structures from the primordial density fluctuations through hierarchical mergers. While $N$-body simulations predict a universal Navarro-Frenk-White mass profile \citep{navarro97}, baryonic physics and its interplay with DM play a fundamental role in re-sculpting DM haloes. Recent comparisons between SL models of massive galaxy clusters and high-resolution cosmological hydrodynamical simulations show that the latter cannot reproduce the observed galaxy-galaxy SL cross-section \citep{meneghetti20,meneghetti22,meneghetti23}, as they predict galaxy cluster sub-haloes to be less compact than what is found in SL models \citep{ragagnin22}. Improved SL modelling techniques can be used to test this discrepancy and potentially shed light on the complex baryon-DM interplay that affects galaxy formation and evolution in clusters, as well as potentially uncover issues in their current description within the $\Lambda$ cold dark matter ($\Lambda$CDM) framework.

In addition, SL models of galaxy clusters have the potential to play a crucial role in cosmographic studies. A $5 \sigma$ discrepancy between the measured values of the Hubble constant, $H_0$, obtained through the study of the anisotropy of the cosmic microwave background %\citep{aghanim20} 
\citep{planck20} 
and the Cepheid distance ladder with type Ia supernovae \citep[SNe;][]{riess22} has recently emerged.
%, and could unveil systematics affecting the measurements \citep{efstathiou20,riess24} or shortcomings in the $\Lambda$CDM paradigm. 
%\citet{Freedman+24} recently found lower $H_0$ from the Tip of the Red Giants and the JAGB stars as calibrators in comparison to the Cepheids, although \citet{Riess+24} showed this is likely due to sample selection.  The $H_0$ could be due to systematic effects such as underestimated dust extinction of SNe Ia \citep[e.g.,][]{Wojtak+22, Wojtak+24}  
The discrepancy is under debate, partly because effects such as the choice of calibrators, sample selection, and dust corrections of SNe Ia could potentially affect $H_0$ measurements \citep[e.g.][]{efstathiou20, riess24, Freedman+24, Riess+24b, Wojtak+22, Wojtak+24}.  Independent methods for assessing the discrepancy are therefore helpful \citep{moresco22} since they would unveil potential systematic effects or new physics beyond the $\Lambda$CDM paradigm.  Quasars and SNe lensed by galaxy clusters (or individual galaxies) can be used as independent precise probes of $H_0$ (i.e. they are not affected by the same systematics and degeneracies) via the measured time delays between multiple images of the variable background source \citep{refsdal64}. These delays are caused by the lens gravitational potential and by different length of the light paths connecting the source to the multiple images. The light curves of SNe are %well known, allowing
drastically changing, which enables measurements of the time delays, typically of the order of weeks to years, with 
%a $1-3\%$ uncertainty \citep{kelly23}. 
uncertainties of less than a few days \citep{kelly23, dhawan20, pierel24a, huber24}.
The time delays depend on the exact value of the Hubble constant and on the total gravitational potential of the cluster; an accurate SL model of the total mass distribution in the cluster core is thus key to obtaining a reliable $H_0$ measurement from the time delays. 

\begin{figure*}
   \centering
   \includegraphics[width=\textwidth]{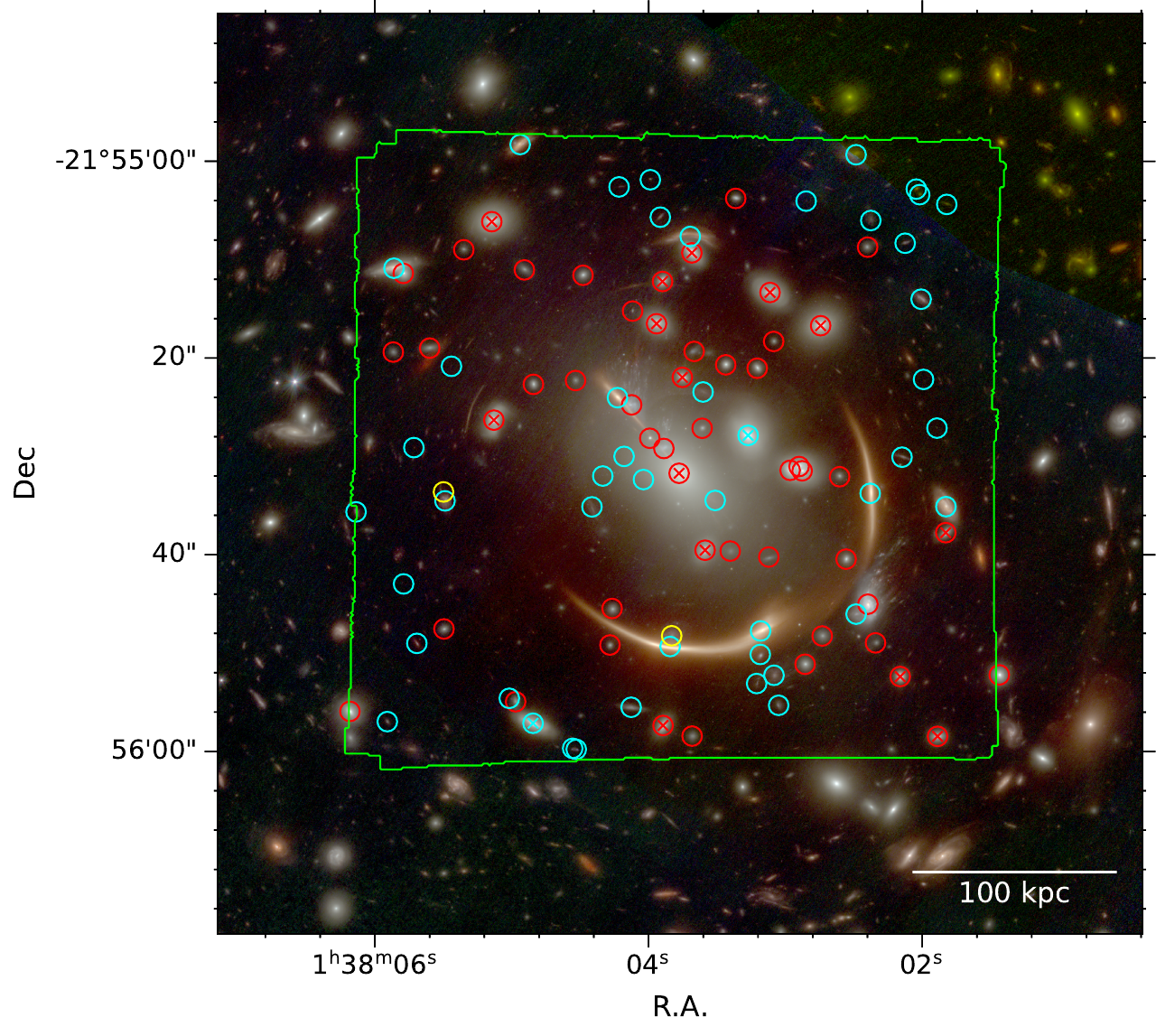}
      \caption{MUSE FoV of MACS0138 overlaid on a colour image built from JWST NIRCam images (red: F277W, F356W, F444W; green: F150W, F200W; blue: F115W). The green region shows the FoV of the final MUSE data cube combining all exposures. Circles indicate spectroscopically confirmed foreground galaxies (yellow), cluster members (red), and background objects (cyan). We mark the cluster members and background galaxies for which we present a stellar velocity dispersion measurement with crosses.}
         \label{fig:M0138_muse_fov}
   \end{figure*}

Improving the accuracy of our description of the total mass distribution of galaxy clusters on sub-halo scales is thus crucial, both for probing the properties of the cluster sub-structures and measuring $H_0$ using time-variable sources. Constraining the mass structure of the cluster members with only cluster-scale SL modelling, however, potentially leads to parametric degeneracy that hinders the reconstruction of their compactness \citep{granata23}. Several recent works have shown how this degeneracy can be broken by taking advantage of observations from Multi Unit Spectroscopic Explorer \citep[MUSE;][]{bacon12} at the Very Large Telescope (VLT). The $1 \, \mathrm{arcmin}^2$ field of view (FoV) of MUSE covers the typical size of the central region of massive clusters at $z\approx 0.3$ (see Fig. \ref{fig:M0138_muse_fov}), allowing one to securely identify cluster galaxies \citep[e.g.][]{mercurio21,lagattuta22} and to obtain priors on the total mass distribution of the cluster members from their measured stellar kinematics \citep{bergamini19,granata22,beauchesne24}.  Furthermore, MUSE allows us to build a complete spectroscopic catalogue that can be used for the detection of an unprecedented number of spectroscopically confirmed lensed background sources within an extended redshift range \citep[up to $z \sim  6$; e.g.][]{richard15,grillo16,jauzac19,bergamini23,caminha23}, which provide very stringent constraints on the lens total mass distribution and help us break the mass-sheet degeneracy affecting its reconstruction.

This work focuses on the MUSE observations of the galaxy cluster MACS J0138$-$2155 \citep[hereafter MACS0138;][]{newman18a,newman18b}. After the observation of the strongly lensed SN Requiem, a likely Type Ia SN \citep{rodney21}, in a host galaxy at redshift $z\approx2$, a second SN, SN Encore, suitable for time-delay cosmography, was observed in the same host galaxy \citep{pierel24b}. The SN was spectroscopically studied in \cite{dhawan24} and identified as Type Ia. In this paper we present a new MUSE spectroscopic catalogue of MACS0138 that contains 107 reliable redshift measurements, including 50 galaxy cluster members and 13 lensed images from four background sources, which we used to build a set of new, independent SL models of the galaxy cluster based on a reliable selection of the cluster members and a pure sample of multiple images. We studied the stellar kinematics of 16 galaxies (14 cluster members and 2 background objects) falling within the FoV of MUSE, and calibrated the Faber-Jackson relation \citep{faber76} for the cluster members, which is crucial for breaking the degeneracy between the values of the velocity dispersion and the truncation radius affecting the total mass reconstruction of cluster members in SL models \citep{bergamini19}.

% Throughout the paper, we adopt a flat $\Lambda$CDM cosmology with  and 
To get an indication of the physical scales from the observed angular scales of our system, we used a flat $\Lambda$CDM cosmology with $H_0=70\, \rm{\, km\, s^{-1}\, Mpc^{-1}}$ and $\Omega_{\rm M} = 1 - \Omega_{\Lambda}=0.3$, where $1''$ corresponds to a scale of 4.809 kpc at $z=0.336$, the redshift of MACS0138. All magnitudes are expressed in the AB system.  
% \sherry{Where is cosmology needed for the analysis?  Conversion of velocity dispersion to Einstein radius? This is independent of H0.  Let's clarify what the adopted flat LCDM model is used for.  Otherwise sounds circular to assume a cosmology to then measure cosmology from delays.}

\section{MUSE observations and data reduction}\label{sec:obs}

The galaxy cluster MACS0138 was targeted by two MUSE programmes between 2019 and 2023. In September 2019, the ESO programme ID 0103.A-0777 (P.I. A. Edge) obtained data covering the core of the cluster with an exposure time of approximately $49$~minutes on target. Later on, in December 2023, after the detection of SN Encore, the Target of Opportunity programme ID 110.23PS (P.I. S.H. Suyu) obtained 2.9 hours of additional data for the same region of the cluster, to increase the depth of the observations. This dataset was obtained using the ground layer adaptive optics (GLAO) system in order to provide improved spatial resolution due to the reduced point spread function (PSF). The FoV of the combined MUSE observations is shown in Fig. \ref{fig:M0138_muse_fov}. It covers the cluster core with an extension of  approximately $1 \, \mathrm{arcmin}$ across, or $289 \, \mathrm{kpc}$ at $z=0.336$.

We used the MUSE data-reduction pipeline version \texttt{v2.8.5} \citep{weilbacher20} to process and calibrate the two datasets and combine them into one final data cube. We followed the standard procedure, such as bias subtraction, flat-fielding, wavelength calibration, and exposure map alignment. Moreover, we used the self-calibration method, implemented in the reduction pipeline, to minimise the instrumental variations across each integral field unit slice and to improve the background subtraction. We also made use of the Zurich Atmosphere Purge \citep[ZAP;][]{soto16} to remove instrumental and sky residuals not fully corrected for by the standard reduction recipes. No bright stars for a precise PSF measurement are present within the FoV of approximately $1 \, \rm arcmin^2$. However, the mean seeing from the Differential Image Motion Monitor station was approximately $0.8''$ during all observations. Moreover, because of the GLAO laser guiding system of the adaptive optics mode used during the observations in December 2023, the region within the wavelength range between $\rm 5800 \, \AA$ and $\rm 5965\, \AA$ was masked out. The final data cube consequently has a reduced effective exposure time in this specific wavelength range, which does not however affect the final spectroscopic measurements. The MUSE data cube was aligned with the \textit{James Webb} Space Telescope (JWST) images from \citet{pierel24b}, using as references the Near Infrared Camera (NIRCam) filter F115W band and the MUSE ‘white image’. We degraded the JWST image to the MUSE white image resolution and employed the positions of approximately $50$ compact sources to match both datasets.

\section{Spectroscopic redshift measurements}\label{sec:zcat}

In order to obtain redshift measurements for as many sources as possible, we followed the method described in Sect. 2.2 of \citet{caminha23}. We first extracted the spectra of all JWST detections in order to measure their redshifts. In this step, we adopted circular apertures with a $0.8''$ radius to obtain one-dimensional spectra of all objects. We carefully inspected all spectra, and in the cases with continuum detection we cross-correlated the data with templates in order to obtain precise redshift measurements. With the help of spectral template matching of different galaxies, as well as the identification of emission lines (such as [OII], Balmer lines, Lyman-$\alpha$, and UV carbon lines), we built our redshift catalogue. We assigned a quality flag (QF) to each redshift measurement, following \cite{balestra16} and \cite{caminha16}, which quantifies its reliability: ‘insecure’ (QF = 1), ‘likely’ (QF = 2), ‘secure’ (QF = 3), or ‘based on a single emission line’ (QF = 9).  The QF = 9 can be considered as secure since the MUSE spectral resolution allows us to distinguish the shape or doublet nature of narrow emission lines (e.g. Lyman-$\alpha$ and [OII]). In a second step, we performed a blind search in the MUSE data cube for emission lines that are not associated with a photometric detection.
For instance, many Lyman-$\alpha$ emitters have no optical or near-infrared photometric counterpart \citep[see e.g.][]{delavieuville19,wisotzki18,claeyssens22,thai23}.
This search is done using a combination of automatic detection on narrow-band images built from the data cube and a visual inspection. In total, 17 emitters (mainly Lyman-$\alpha$) are detected in this step.

The full spectroscopic sample contains 107 reliable (i.e. likely or secure, $\rm QF \ge 2$) redshift measurements of extragalactic objects, two are foreground ($z < 0.324$) objects, 58 are at the cluster redshift ($0.324 < z < 0.349$; see Sect. \ref{sec:cmemb}), including 50 cluster member galaxies, 47 redshift measurements are in the cluster background ($z > 0.349$; see Sect. \ref{sec:back}), out of which we identify 13 lensed multiple images from four background emitters. The redshift catalogue is summarised in Table \ref{tab:z_muse}. 
% \sherry{Reminder: double check consistency in numbers with Ertl et al.~paper, especially in final phase in case there are revisions}  

\subsection{Cluster galaxies}\label{sec:cmemb}

\begin{figure}
        \centering
        \includegraphics[width = 1.0\columnwidth]{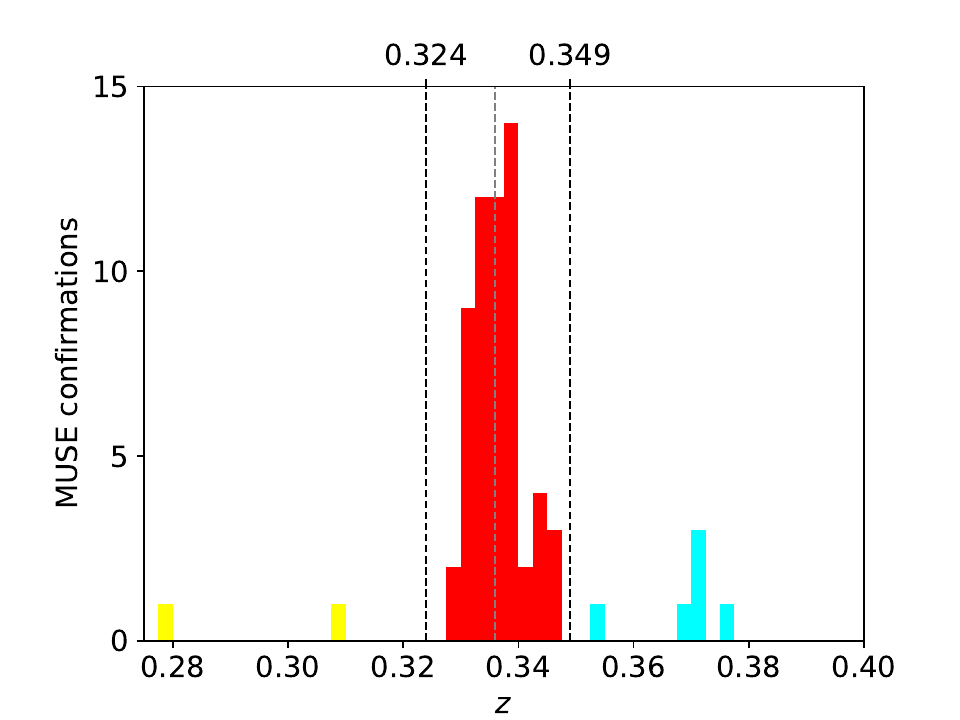}
        \caption{Redshift distribution around the cluster redshift ($z_{\rm cluster}=0.336$, marked with a dashed grey line). Cluster members are indicated in red, and foreground and background objects in yellow and cyan, respectively. The vertical lines indicate the region with a relative velocity of $\rm \pm 2500~km~s^{-1}$ from the cluster mean redshift.}
        \label{fig:cluster_hist}
\end{figure}

The cluster MACS0138 is clearly visible in the distribution of the redshift measurements included in Table \ref{tab:z_muse} as an over-density around the redshift range of $0.32 \lesssim z \lesssim 0.35$. We identified the spectroscopic cluster members by selecting all galaxies with a reliable redshift estimate (i.e. with a $\mathrm{QF} \ge 2$) and a rest-frame relative line-of-sight velocity within $\Delta V = 2500 \, \mathrm{km\,s^{-1}}$ with respect to the cluster mean line-of-sight velocity, determined from the width of the redshift space cluster member distribution, which corresponds to the redshift range $z=0.324-0.349$. As clear from Fig. \ref{fig:cluster_hist}, the redshift distribution is well centred around $z_{\rm cluster}=0.336$. Our catalogue contains 58 reliable redshift measurements in the relevant redshift range. This spectroscopic sample was used by \cite{ertl25}, \cite{acebron25}, and Suyu et al. (in prep.) as a basis for the catalogue of cluster members included within the strong lensing models presented therein. They excluded eight non-galactic objects from the catalogue, typically star forming clumps, marked in Table \ref{tab:z_muse} with an asterisk, for a final catalogue comprising 50 spectroscopically confirmed cluster members. A visual spectral study of the 50 cluster members reveals that the vast majority of the sample (45 members), as expected, shows spectra typical of old, passively evolving stellar populations. This includes the brightest cluster galaxy (BCG), whose spectrum is strongly contaminated by broad emission lines, presumably from its active galactic nucleus. The remaining five cluster members are active (mostly blue or jellyfish galaxies). In Sect. \ref{sec:kin} we use the MUSE spectra to probe the stellar kinematics of the brightest red members, in order to obtain a prior on their total mass structure. The cluster member catalogue was completed in \cite{ertl25} with the addition of non-spectroscopic members selected on the basis of the colour-magnitude and colour-colour relations of the cluster.   

\subsection{Background objects and multiple image systems}\label{sec:back}

\begin{figure*}
        \centering
            \includegraphics[width = \textwidth]{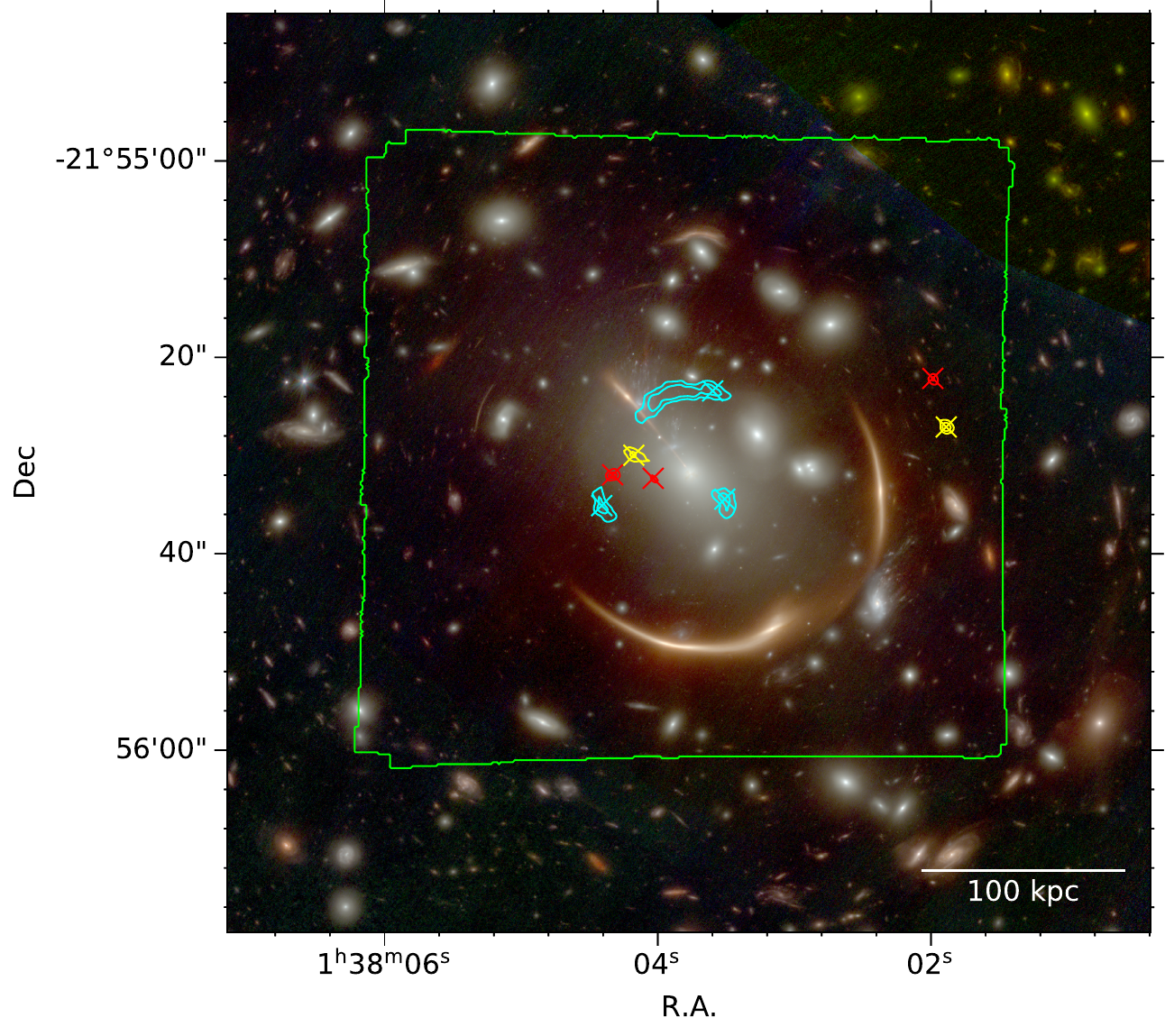}
        \caption{MUSE FoV of MACS0138 overlaid on a colour image built from the JWST NIRCam images (the bands used are the same as in Fig. \ref{fig:M0138_muse_fov}). The cyan, red, and yellow lines show iso-surface-brightness contours for the MUSE cube at the wavelengths of the emission lines used to detect the %three newly-identified 
        three lensed sources: [OII] emission at $z=0.767$ (cyan), Lyman-$\alpha$ emission at $z=3.152$ (red) and Lyman-$\alpha$ emission at $z=3.420$ (yellow).  The two Lyman-$\alpha$ emissions are newly discovered sources strongly lensed by this cluster.
        %. In cyan, [OII] emission at $z=0.767$, in red and in yellow Lyman-$\alpha$ emission at $z=3.15$ and $3.42$, respectively. 
        We mark the position of the secure multiply lensed images detections ($\mathrm{QF}=3$) for the three sources (Systems %4/5, 6, and 7,
        4, 5, and 6) with a cross.}
        \label{fig:muse_sources}
\end{figure*}

\begin{figure*}[t!]
  \includegraphics[width = 0.68\columnwidth]{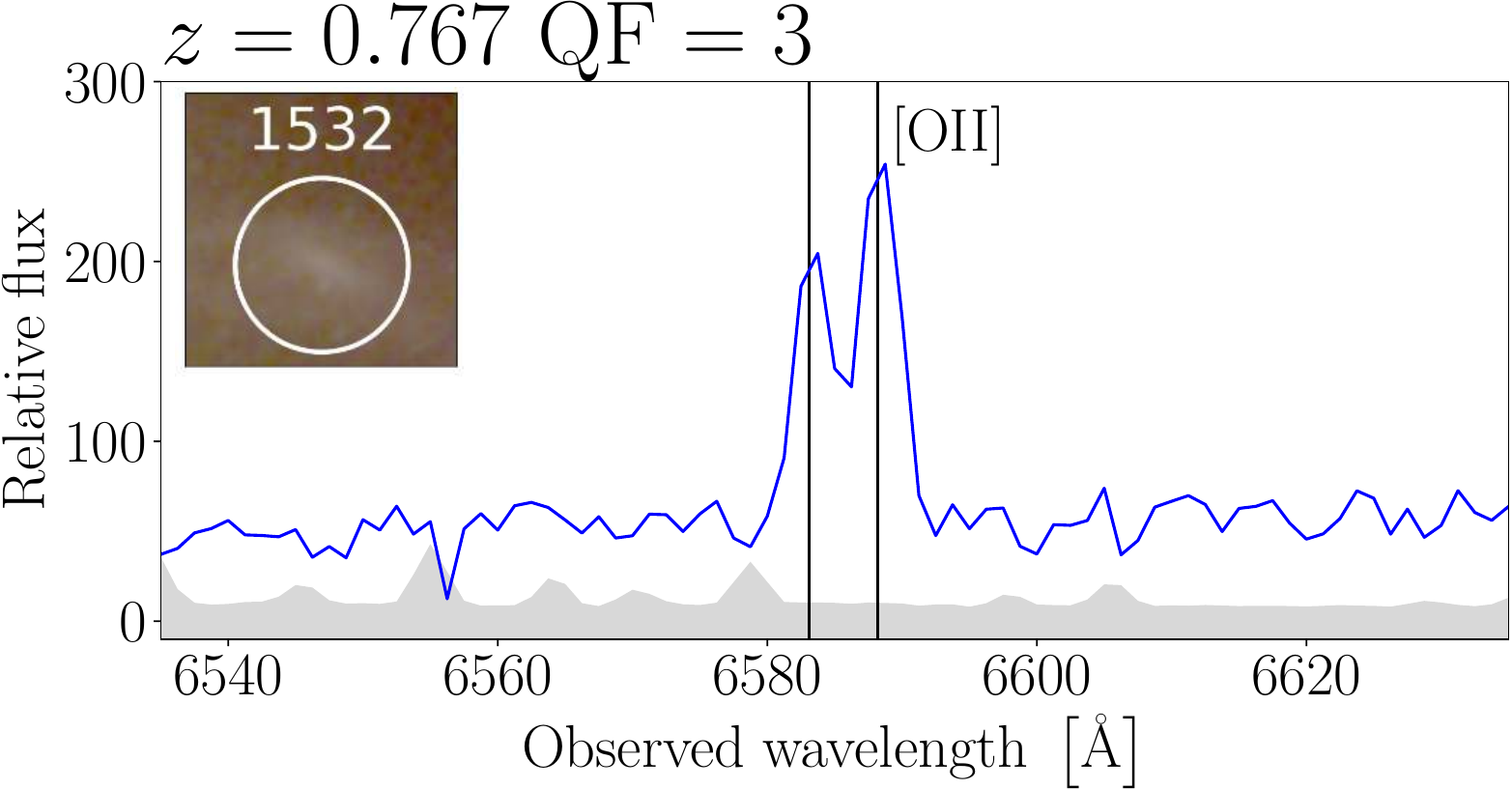}
  \includegraphics[width = 0.68\columnwidth]{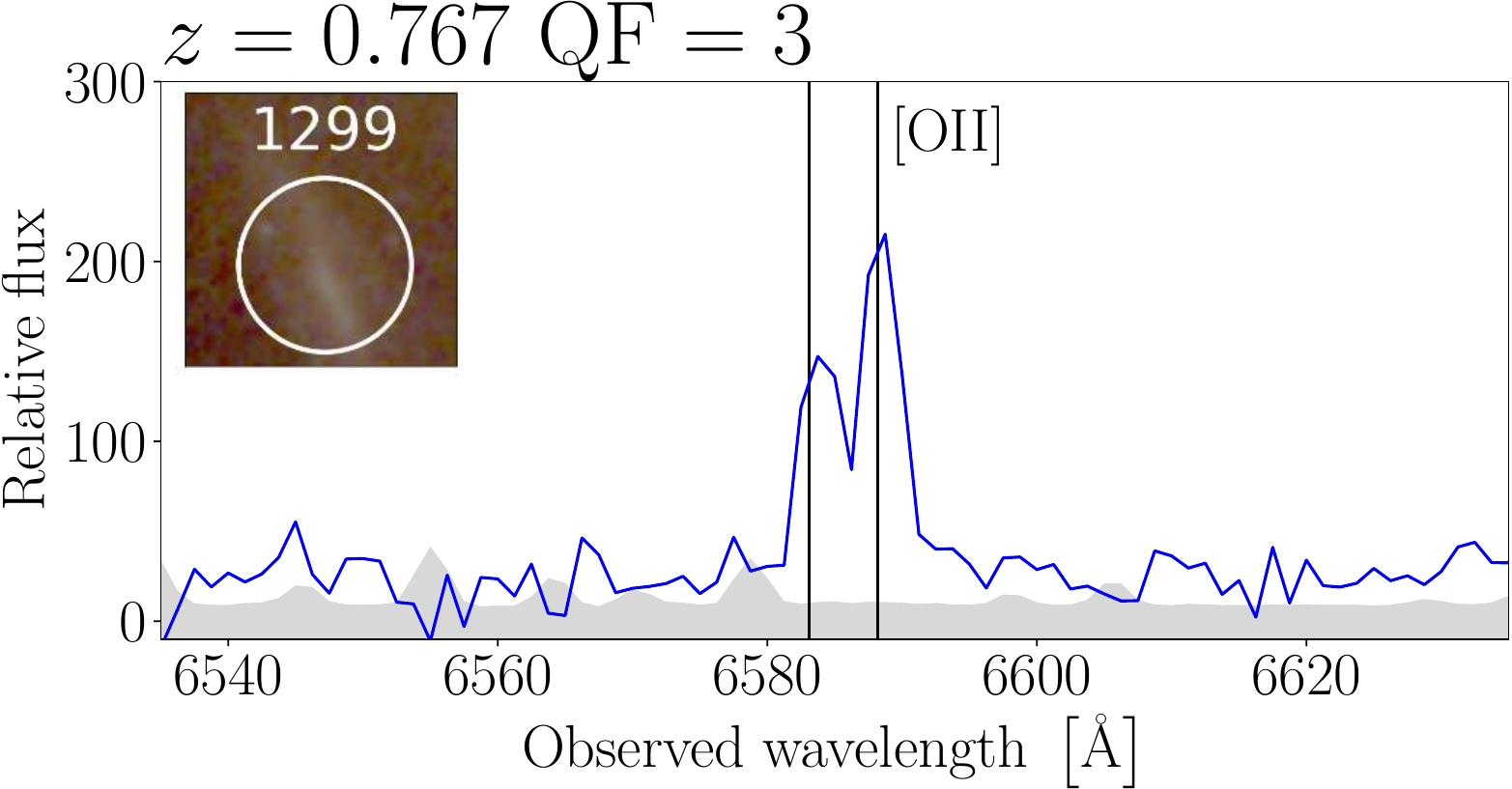}
  \includegraphics[width = 0.68\columnwidth]{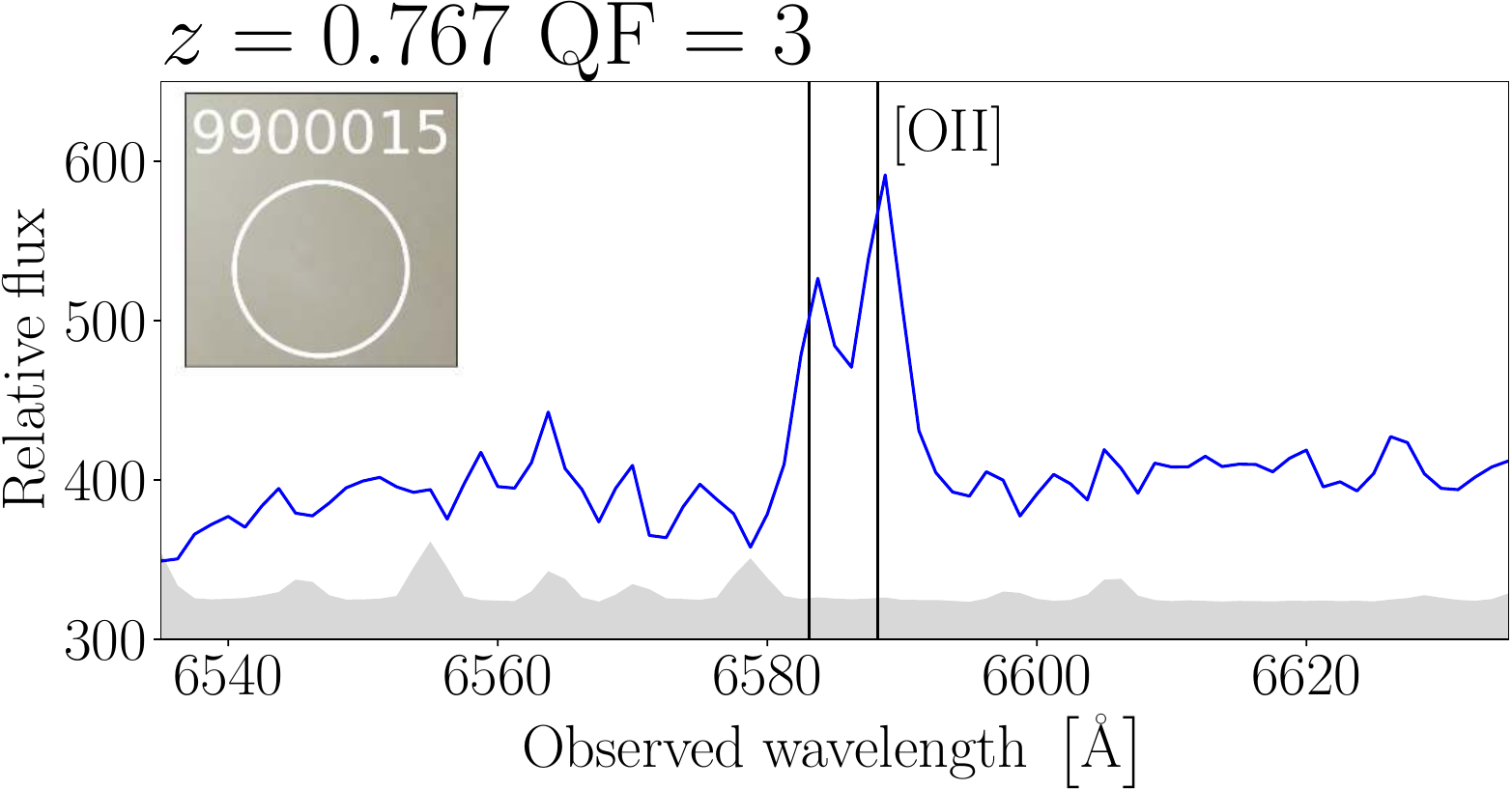}
  \includegraphics[width = 0.68\columnwidth]{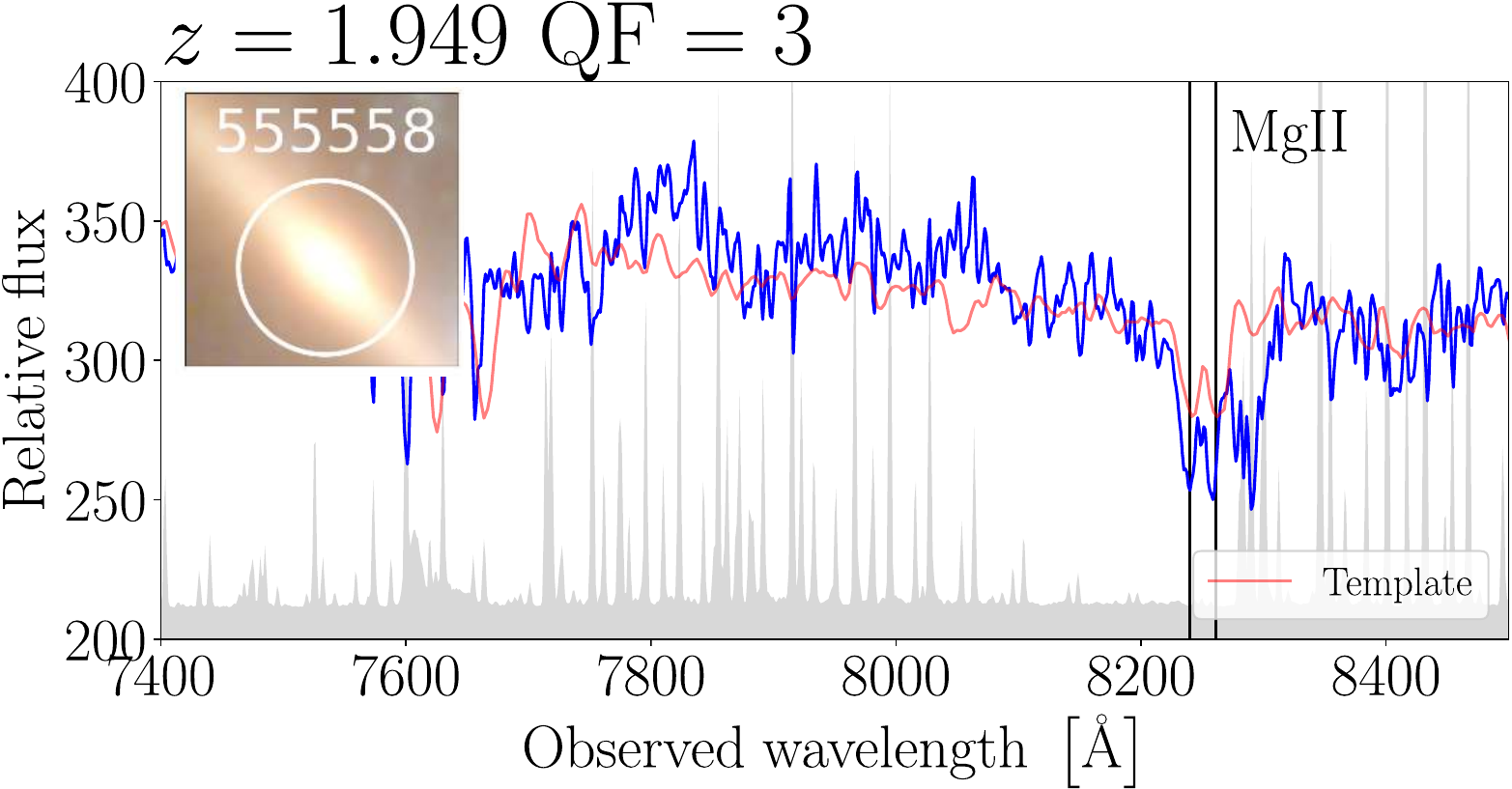}
  \includegraphics[width = 0.68\columnwidth]{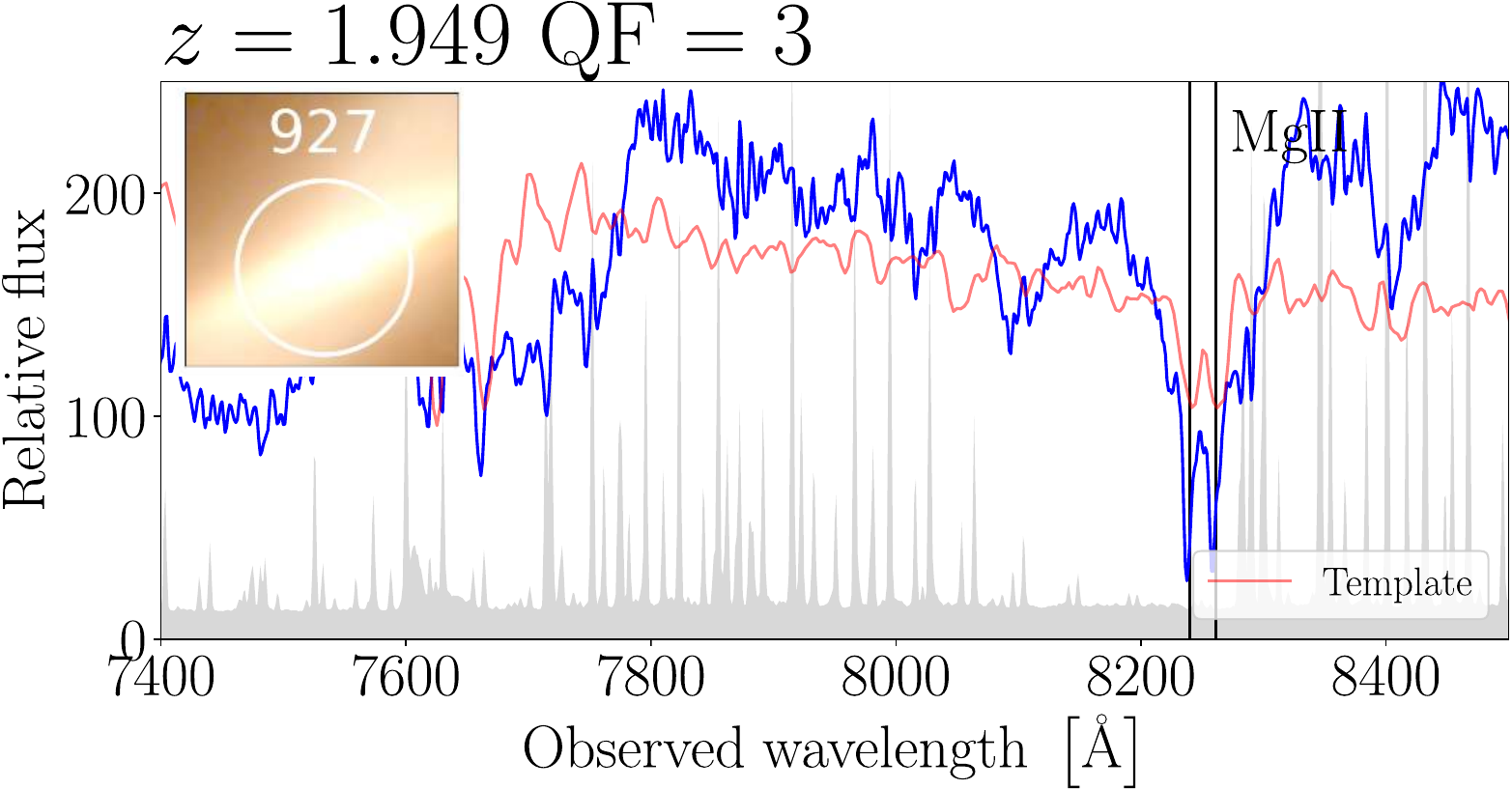}
  \includegraphics[width = 0.68\columnwidth]{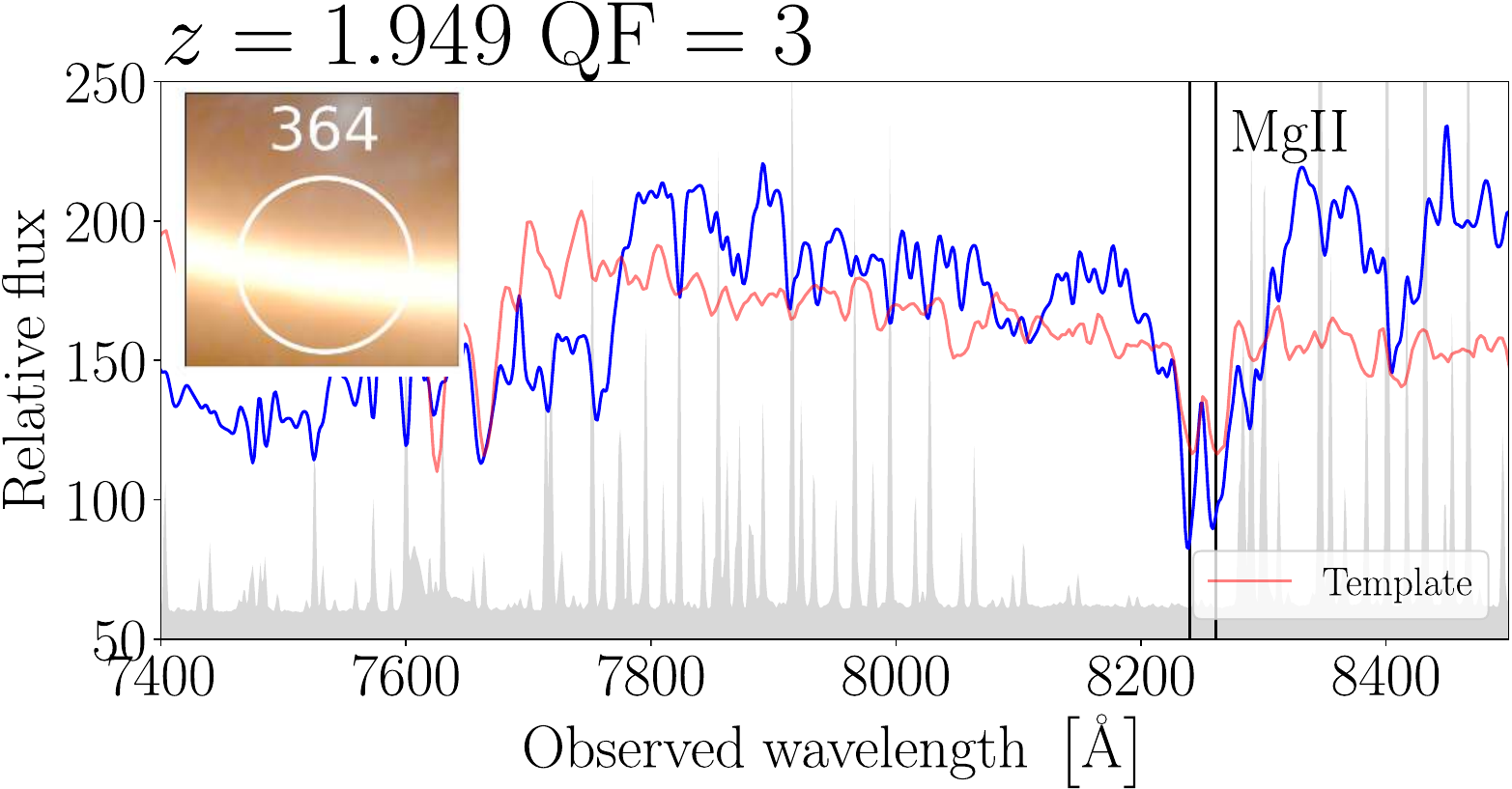}
  \includegraphics[width = 0.68\columnwidth]{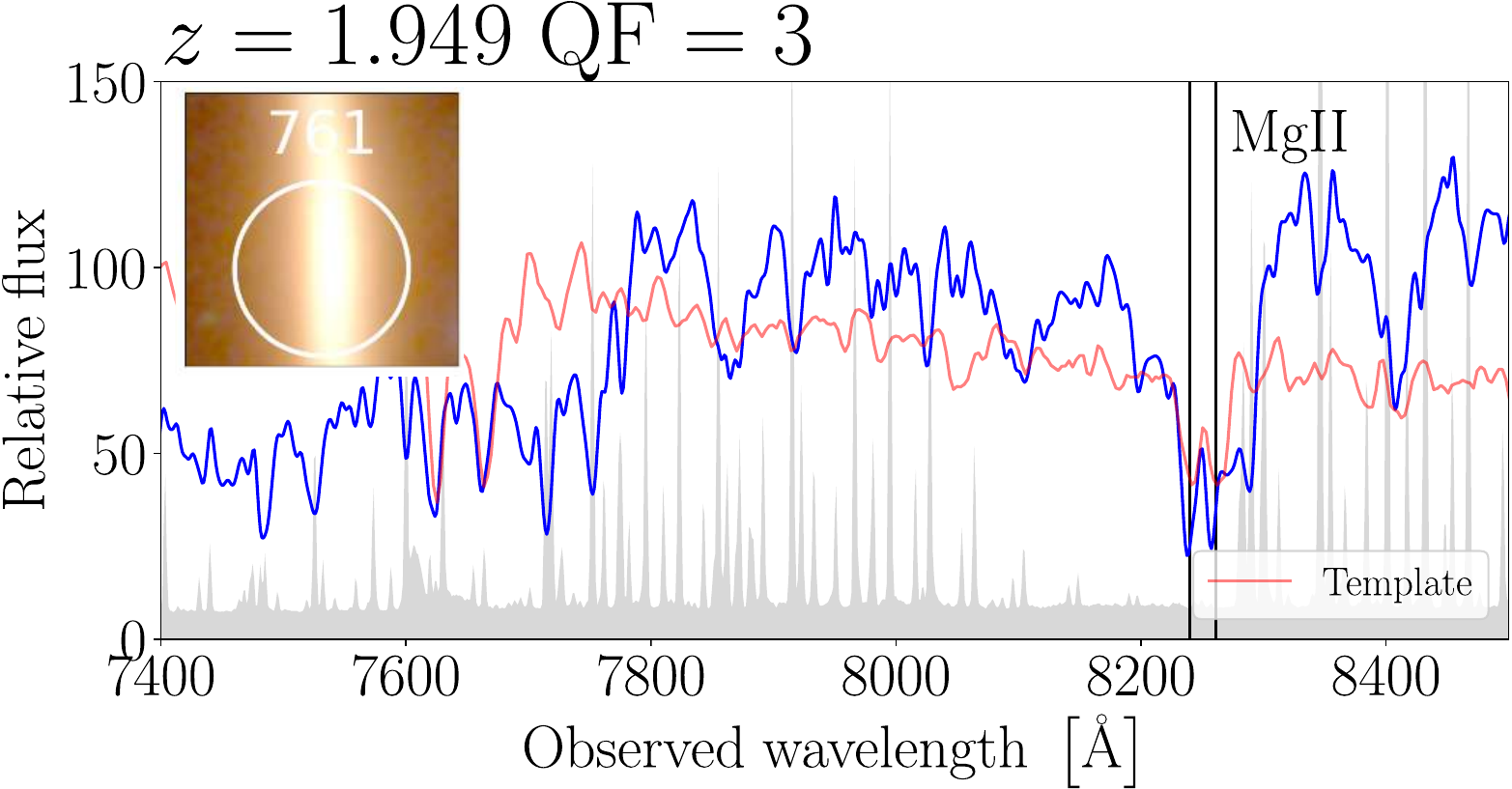}
  \includegraphics[width = 0.68\columnwidth]{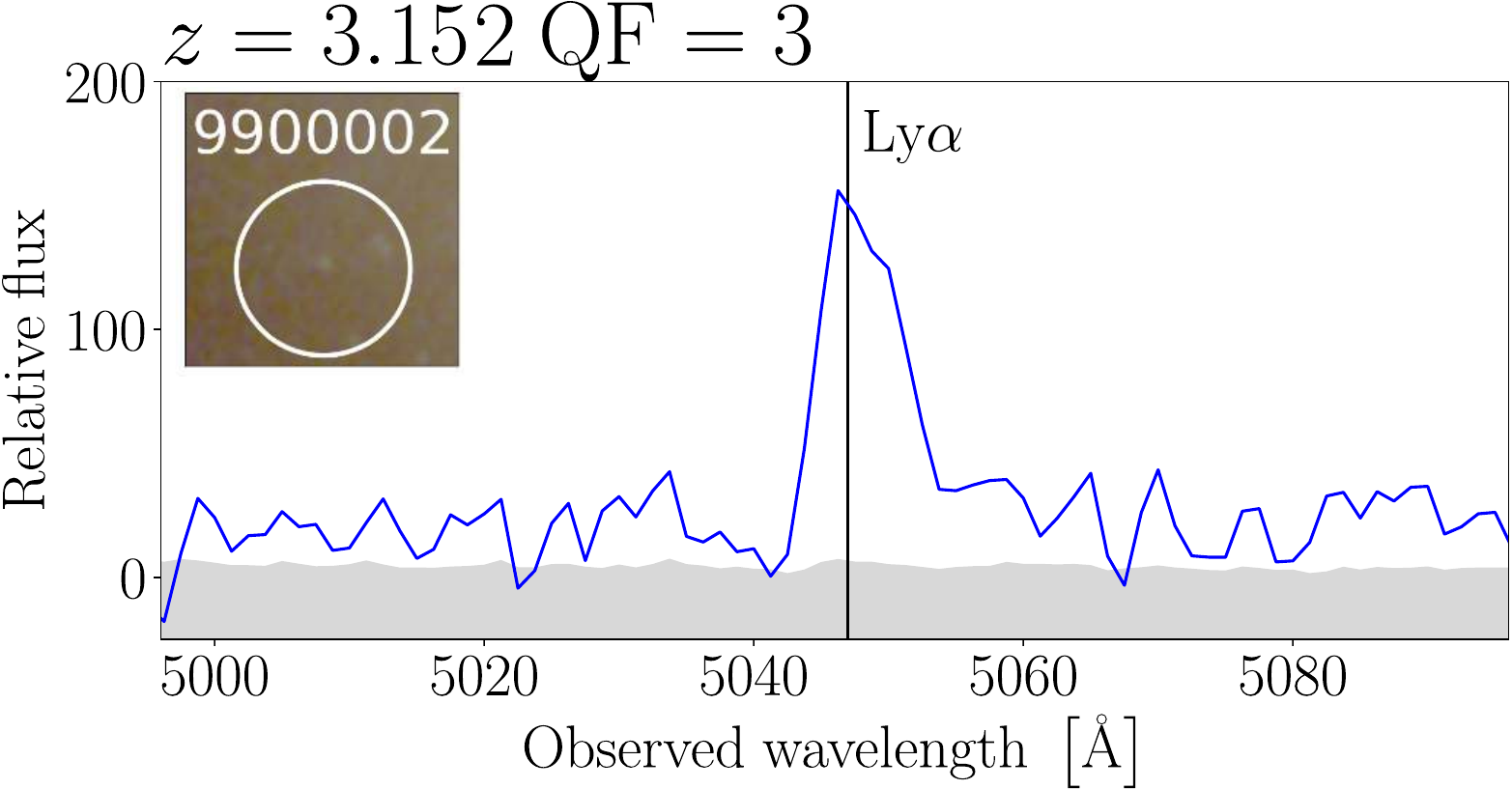}
  \includegraphics[width = 0.68\columnwidth]{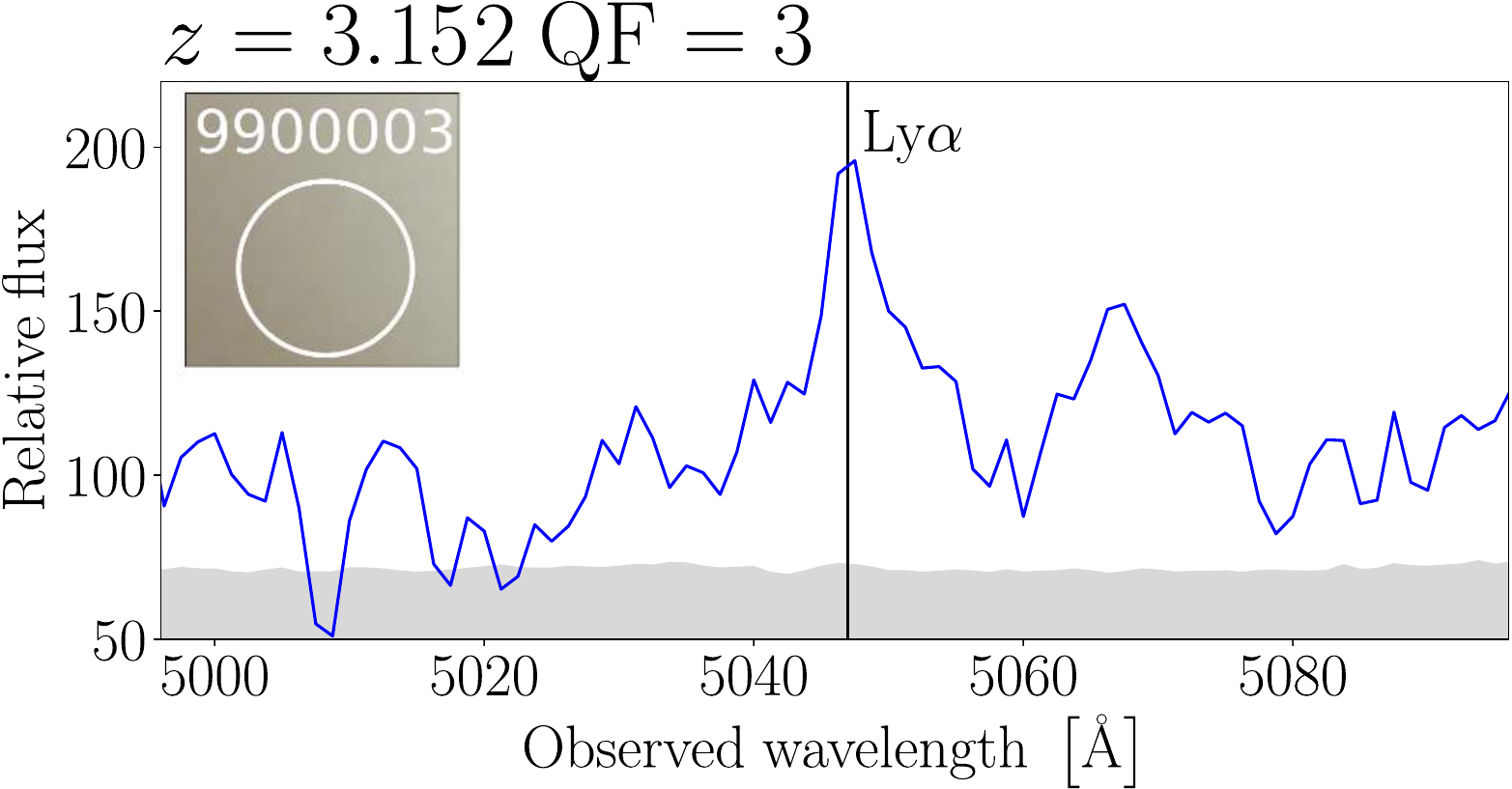}
  \includegraphics[width = 0.68\columnwidth]{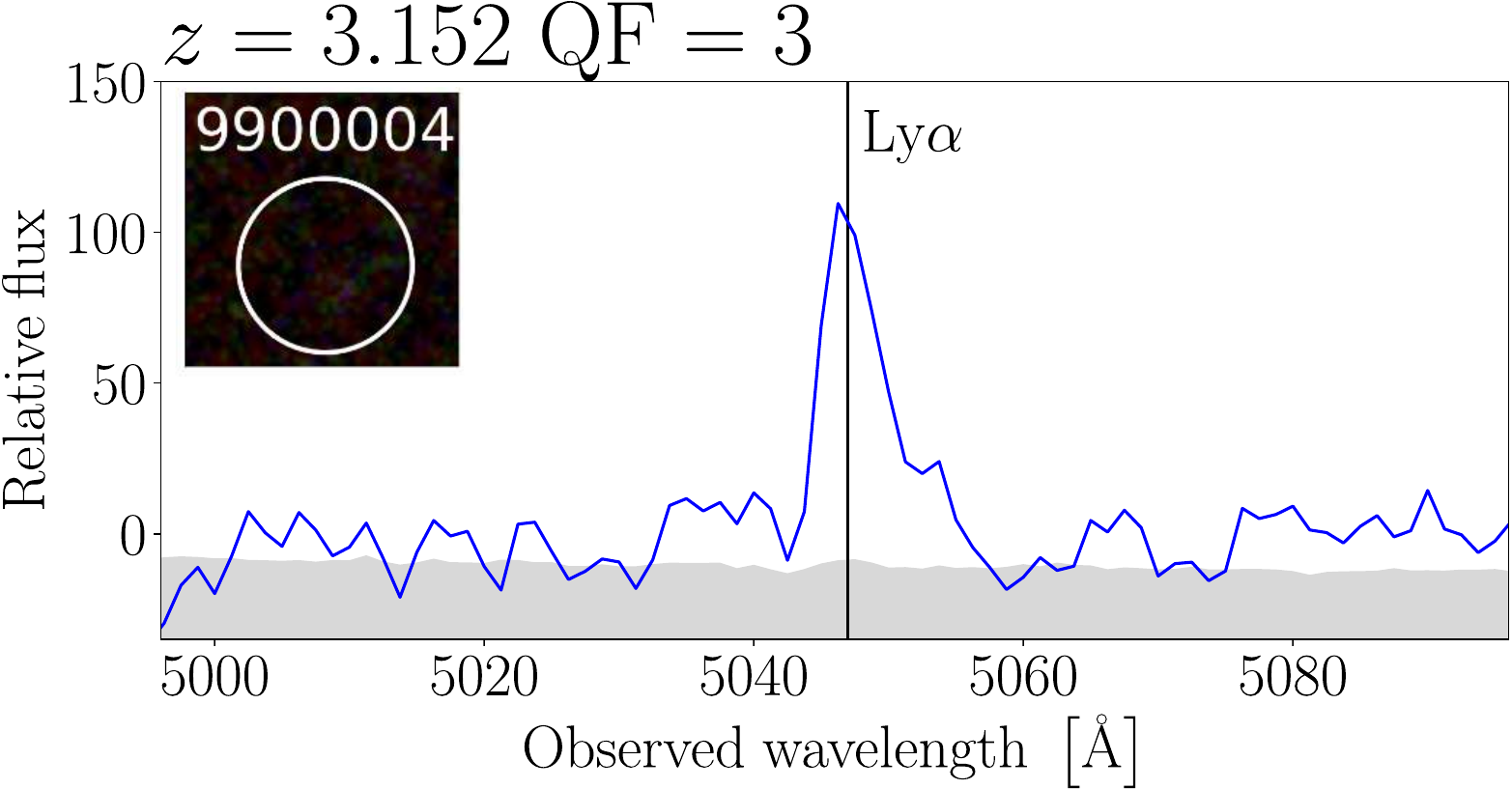}
  \includegraphics[width = 0.68\columnwidth]{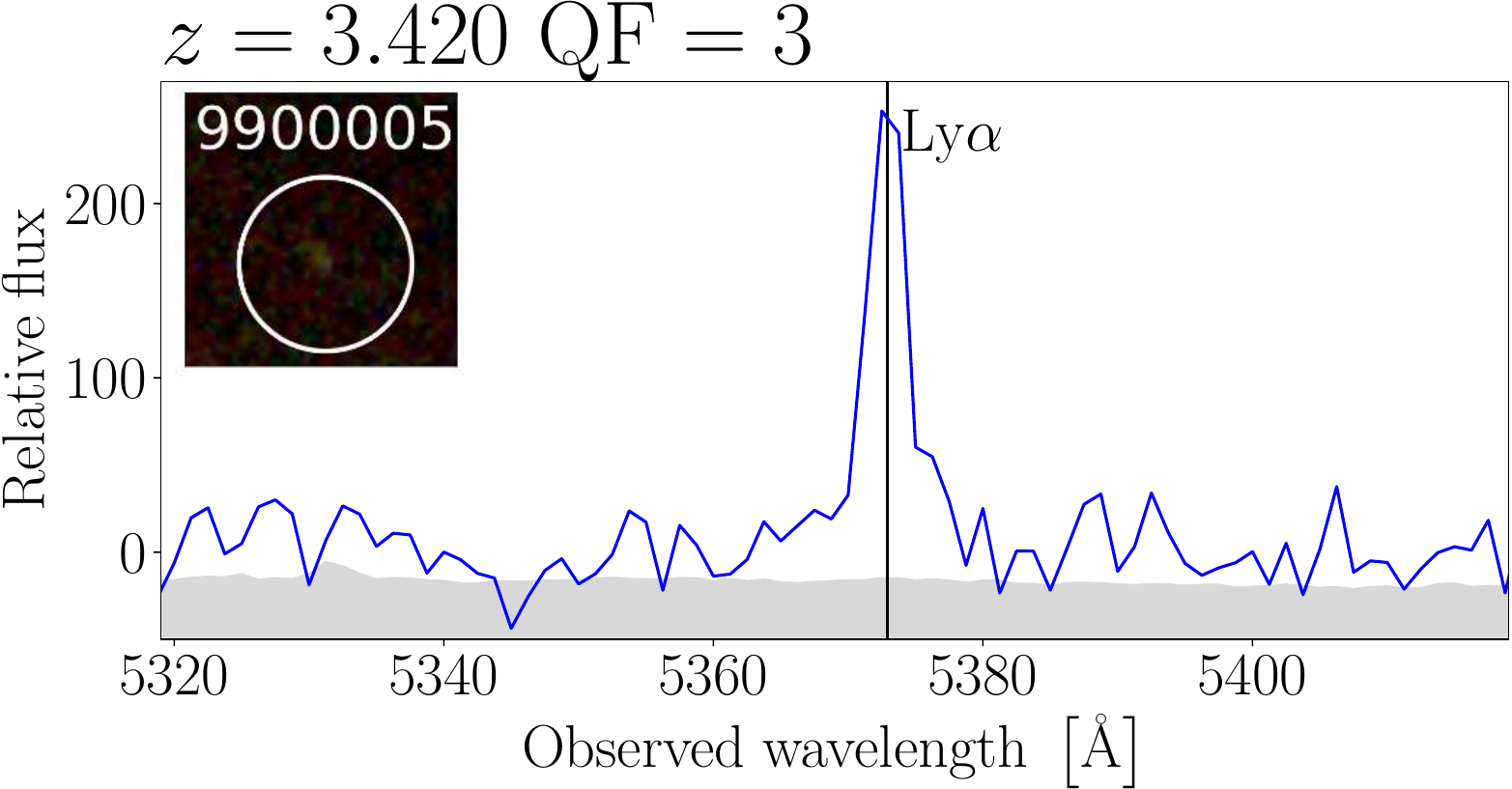}
  \includegraphics[width = 0.68\columnwidth]{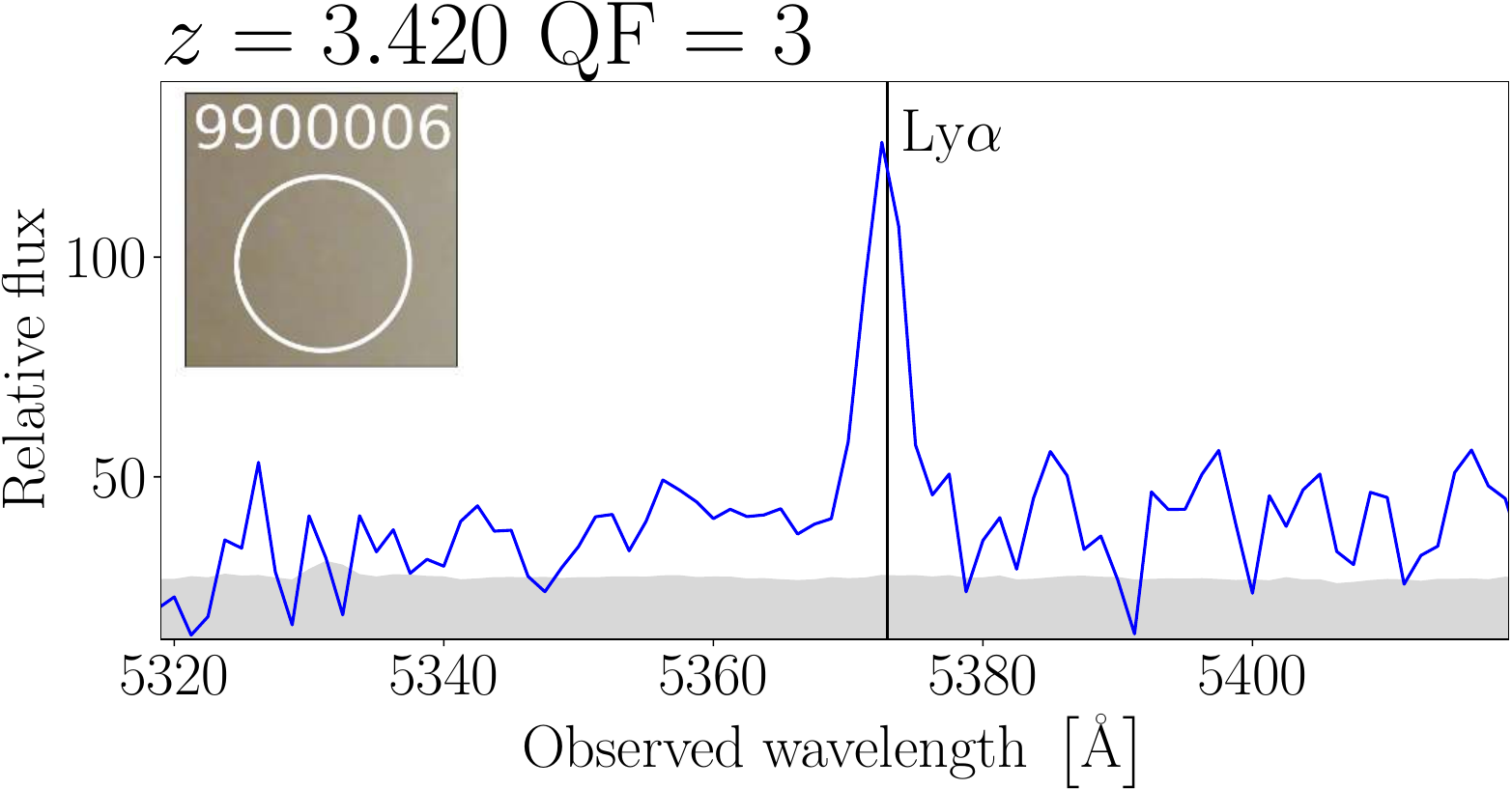}
  \caption{Spectra of the secure multiple images ($\rm QF =3$) included in our catalogue. The vertical line indicates the observed position of the main spectral feature used for the identification ([OII], MgII, and Lyman-$\alpha$). We show the re-scaled variance in grey. The measured redshift and its QF are indicated in each figure title. The HST cutouts are $2\arcsec$ on each side, and we provide the ID\textsubscript{MUSE} of the object in Table \ref{tab:z_muse}.}
  \label{fig:multiple_image_spectra}
\end{figure*}

The catalogue contains 47 redshift measurements with $z>0.349$, from 38 individual background emitters, when accounting for multiple images of the same sources. The redshift space distribution shows a secondary peak at $z \approx 0.37$, visible in Fig. \ref{fig:cluster_hist}, possibly from a smaller group of galaxies in the background. Their impact on the strong lensing model is accounted for in \cite{ertl25}, \cite{acebron25}, and Suyu et al. (in prep.). The catalogue contains four lensed images of the host of SN Encore and SN Requiem, at $z=1.95$, confirmed from the MgII absorption doublet. A fifth image was identified in \cite{ertl25}, blended with the BCG light. In addition to the three multiple images from an [OII] emitter that were previously reported in \citet{rodney21}, we were able to identify two further lensed background sources at redshifts $z = 3.152$ and $z = 3.420$, for a total of 13 lensed multiple images with a reliable redshift measurement (i.e. QF$\ge 2$), which we mark in bold in Table \ref{tab:z_muse}. The search was performed both from our spectroscopic catalogue and through a blind search for simultaneously appearing spectral features when scanning the MUSE data cube. Our multiple image catalogue was used as the basis for the new SL models. Specifically, we identify 
%triple 
multiple lensed images of a star-forming galaxy and two high-$z$ Lyman-$\alpha$ emitters. The first source is at $z=0.767$, with prominent [OII] and [OIII] emission lines, including a faint image appearing close to the centre of the BCG. Owing to its clumpy nature, two components of the source have been identified in \cite{ertl25} and their multiple images have been used as separate constraints for the SL model (System 4). The two Lyman-$\alpha$ sources are at $z=3.152$ and $3.420$ and were identified from scanning the MUSE cube, as they lack a clear photometric counterpart in \textit{Hubble} Space Telescope (HST) and JWST data. They are included in the SL model by \cite{ertl25} as Systems 5 and 6. The position of the emission line sources can also be inferred on the basis of the MUSE surface brightness at the corresponding wavelength, which we show in Fig.~\ref{fig:muse_sources} for the newly identified lensed systems. In Fig. %\ref{fig:muse_sources}, 
\ref{fig:multiple_image_spectra},
instead, we show the MUSE spectra of the secure ($\rm QF =3$), marking the spectral emission or absorption features used for their identification and confirmation. Finally, we identify an arc-like region (ID\textsubscript{MUSE}=1755) lensed by a cluster member (ID\textsubscript{MUSE}=1631). We extracted its spectrum from an arc-shaped aperture, to avoid contamination from the nearby member, and measured a likely (QF=2) redshift $z=1.945$. Given that its value is not secure, the redshift is not used as a constraint 
%, and left free to vary, for the optimisation of the SL models.
in the `gold' sample of lensed images for SL models.

% In addition, we also obtained a tentative redshift for a galaxy scale system with possible $z=2.7745$ \granata{remove? Should we add the new silver system by the Ferrara team?}. In Table ??, we provide the coordinates and MUSE redshift values of all multiple images used as input for the lens model \granata{This is in Ertl, remove?}.  and their distribution is illustrated in Fig. 

\section{Stellar kinematics of galaxies}\label{sec:kin}

The depth of the available MUSE data enables an accurate study of the line-of-sight stellar kinematics of several member and line-of-sight galaxies, providing us with a further tracer of their total mass distribution, which has been used in \cite{ertl25} to enhance the accuracy of the description of the galaxy-scale mass components within the strong lensing models of MACS0138. In this section we briefly describe the pipeline for the spectral measurement of the line-of-sight stellar velocity dispersion of the bright galaxies falling within the MUSE FoV, detailed in Granata et al. (in prep.), and present the kinematic catalogue adopted by \cite{ertl25}. Finally, we present some insights into the total mass structure of the cluster members obtained enhancing strong lensing modelling with priors from stellar kinematics.

\subsection{Measured velocity dispersions from MUSE spectroscopy}\label{sec:kinm}

We measured the value of the line-of-sight stellar velocity dispersion, $\sigma_v$, of each galaxy included in the analysis following the procedure presented in \citet{granata23} and Granata et al. (in prep.). For the kinematic analysis, we used the public spectral fitting code \texttt{pPXF}\footnote{Version 8.2.6} \citep[penalized pixel-fitting;][]{cappellari04,cappellari17,cappellari23} to perform a full-spectrum fit of the continuum and of the absorption lines of the observed spectra in the rest-frame wavelength range $[3700-5100]\, \rm \AA$. The optimisation fits simultaneously for the stellar population and for the line-of-sight stellar velocity distribution of the observed galaxy, by comparing its spectrum with a combination of stellar templates chosen from a set of high-resolution UVB stellar spectra from the X-shooter Spectral Library (XSL) DR3 \citep{verro22}, convolved with a line-of-sight velocity distribution. We based our template selection on the work of \cite{knabel25}, who showed that template selection is the main source of systematics for stellar kinematic studies of early-type galaxies. They examined all 830 XSL DR3 spectra and purged those that are not flux calibrated or corrected for slit loss or extinction. They then visually inspected all remaining templates, flagging any potential issue with them (e.g. problems with extinction or telluric corrections, high noise, or emission lines) and the wavelength range affected. The authors kindly provided us with the list of clean templates and with wavelength flags for possible problems affecting them, allowing us to compile a final library of 462 templates, which we used for our kinematic studies.

We combined the templates with 12th degree additive Fourier polynomials, and we chose a Gaussian line-of-sight velocity distribution, as we are interested in fitting only its first two moments. We checked the robustness of the fit with respect to the choice of the stellar template library, and, as done in \cite{cappellari04}, we verified its reliability with a set of 16,000 simulated MUSE spectra of cluster members, reproducing a variety of spectral type and observing conditions (Granata et al. in prep.). We find that a spectral $S/N\geq10$ is required to guarantee consistently accurate velocity dispersion measurements. Here and in the rest of this paper we use the notation $S/N$ to refer to the average signal-to-noise ratio of the MUSE wavelength bins that we considered for the velocity dispersion fit. Given the spectral sampling of $1.25 \, \rm \AA$ of MUSE, a signal-to-noise per bin of $S/N=10$ corresponds to a signal-to-noise per angstrom of $8 \, \rm \AA^{-1}$. These simulations were also used to obtain a reliable estimate of the error on the value of $\sigma_v$. On one hand, \texttt{pPXF} provides a formal uncertainty on the value of $\sigma_v$, from the covariance matrix of the errors on the fitted parameters, which we re-scaled assuming that the best-fit $\chi^2$ matches the number of degrees of freedom in the fit. Whilst this method can provide an order-of-magnitude estimate of the uncertainty assuming the fit is good, it can significantly under-estimate the uncertainty in the case of noisier spectra or low values of $\sigma_v$ \citep{cappellari04}. In this regime, simulations can yield a more reliable estimate of the accuracy of the recovered values \citep{cappellari04}: in Granata et al. (in prep.) we probe the relative uncertainty on the value of sigma as a function of $S/N$ from the $1\sigma$ scatter of the recovered $\sigma_v$ values about the input velocity dispersion of the 16,000 simulated spectra. Similarly to \cite{bergamini19}, Granata et al. (in prep.) fitted a relation between the relative uncertainty on $\sigma_v$ and $S/N$, which we used here to obtain a more reliable estimate of the error associated with the spectral noise of the recovered $\sigma_v$ value. We chose, conservatively, to sum in quadrature this estimate of the error on $\sigma_v$ with the formal parametric error given by \texttt{pPXF} to obtain our final uncertainty estimates, reported in Table \ref{tab:vdispcat}. 

\begin{figure}
        \centering
        \includegraphics[width = 1.0\columnwidth]{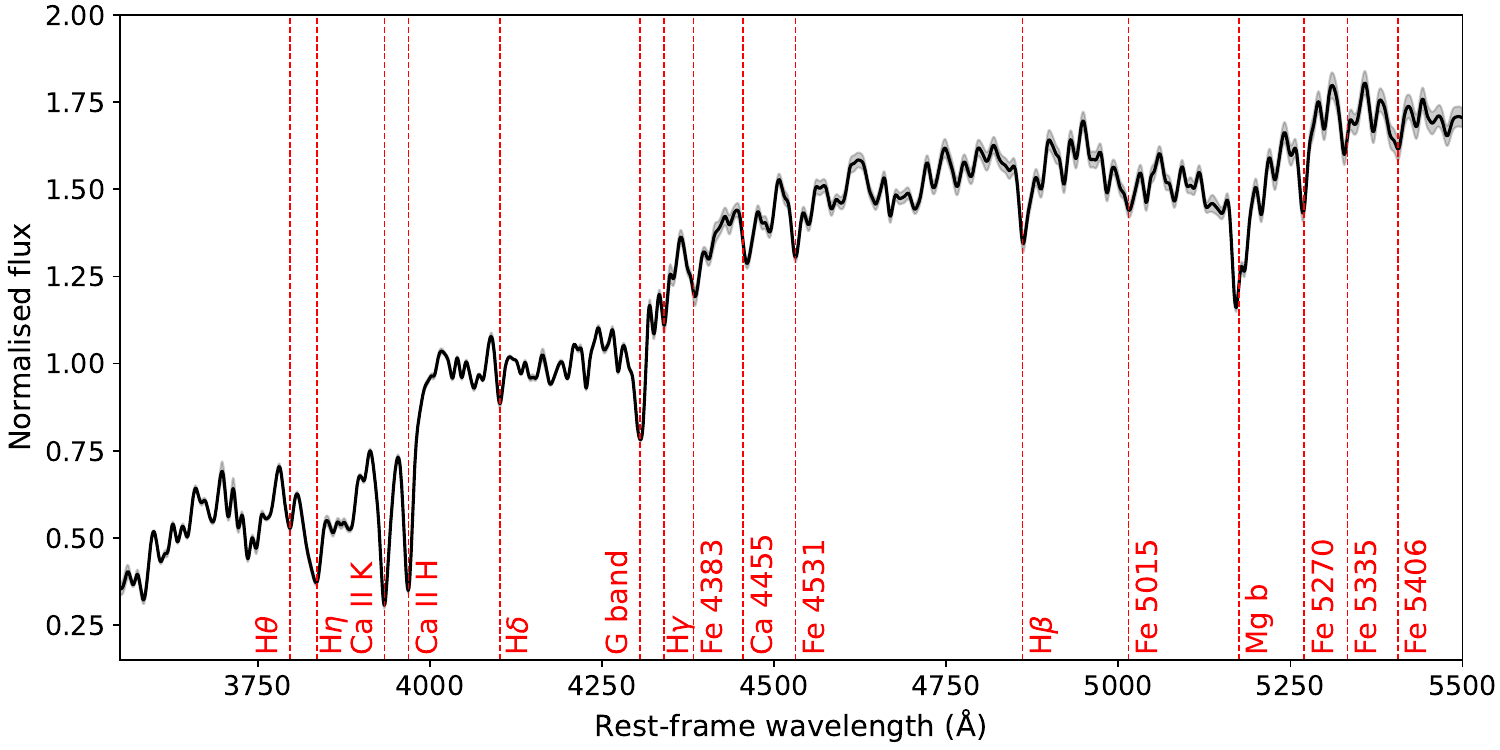}
        \caption{VLT/MUSE mean stacked spectrum of the 13 cluster members, all except the BCG, for which we obtained a line-of-sight stellar velocity dispersion measurement. The shaded region shows the standard deviation of each spectral pixel. The spectra were smoothed by applying a Gaussian kernel with a standard deviation of $3.75 \, \mathrm{\AA}$. The vertical dashed red lines indicate the main absorption features of the spectrum.}
        \label{fig:stack}
\end{figure}

As shown in Granata et al. (in prep.), we can increase the $S/N$ of the spectra, thus extending the sample of cluster members for which we can probe the stellar kinematics, by weighting the MUSE cube with the members' surface brightness. To this purpose, we used the HST F814W band image, de-graded and re-binned to the PSF and pixel-scale of our MUSE observations, and extracted the spectra from large circular apertures with a $1.5''$ radius, centred on the galaxy centre of light. The spectra resulting from the weighted average are representative of the central regions of the members, which are sampled with a higher weight due to their higher surface brightness: the velocity dispersion value obtained from them are on average equivalent to those measured within the galaxy half-light radius. We can thus probe the central velocity dispersion of the cluster galaxies with high $S/N$ and without the need for aperture corrections. In Fig. \ref{fig:stack} we show a VLT/MUSE mean stacked spectrum of the 13 early-type cluster member galaxies (excluding the BCG) for which we obtained a velocity dispersion measurement. The stacked spectrum, obtained from a sum of the spectra of the 13 members normalised using their median flux value in the rest-frame $[4125,4275] \, \mathrm{\AA}$ wavelength range, was smoothed with a Gaussian kernel with a standard deviation of $3.75 \, \mathrm{\AA}$, thus highlighting the spectral features more clearly. The shaded area shows the uncertainty obtained from the standard deviation scatter of the 13 spectra about the mean in each spectral bin. The stacked spectrum is typical of an old, passively evolving, stellar population, and clearly showcases several absorption features, such as the Ca II H and K, the G band, and the H$\beta$ lines, whose width can be used to anchor the fit of the line-of-sight stellar velocity dispersion of the members. 

\begin{table*}
\caption{Catalogue of measured velocity dispersion values.}             
\label{tab:vdispcat}      
\centering          
\begin{tabular}{c c c c c c c}     
\hline\hline              
ID & ID\textsubscript{MUSE} & R.A. & Dec  & $\sigma_v \, (\mathrm{km \, s^{-1}}$) & $\delta \sigma_v \, (\mathrm{km \, s^{-1}})$ & $S/N$ \\  & & (J2000) & (J2000)   & & & \\ 
\hline                    
  55   &   895 & 24.5162254 & $-$21.9326011 & 114.7 & 4.4 &  22.4 \\
  58   &   930 & 24.5078679 & $-$21.9329074 & 97.1 &  5.2 &  18.1 \\
  69   &   1032 & 24.5090052 & $-$21.9312181 & 157.2 & 5.3 &  23.8 \\
  78   &   1127 & 24.5149459 & $-$21.9276509 & 148.0 & 3.9 &  47.1 \\
  102  &   1210 & 24.5213768 & $-$21.9239768 & 176.4 & 3.1 &  51.7 \\
  90   &    1211 & 24.5076178 & $-$21.9271544 & 88.0 & 7.9 &  12.4 \\ 
  115  &   1308 & 24.5114284 & $-$21.9213096 & 210.5 &  3.1 &  56.8 \\
  191  &   1386 & 24.5214401 & $-$21.9183725 & 175.5 & 2.5 &  59.3 \\
  203  &   1433 & 24.5164250 &  $-$21.9212544 & 161.3 &  3.4 &  44.9 \\
  116  &   1457 & 24.5156388 & $-$21.9227696 & 155.1 & 5.0 &  29.7 \\
  154  &   1498 & 24.5129692 & $-$21.9203716 & 137.3 & 2.9 &  43.0 \\
  193  &   1631 & 24.5153499 & $-$21.9192529 & 214.8 & 4.1 &  44.8 \\
  156  &   1633 & 24.5162417 & $-$21.9200615 & 171.0 & 5.2 &  28.5 \\
  $0^*$ &     827 &  24.5157373 & $-$21.9254784 & 321.9 & 16.1 &  52.7 \\
\hline
  51   &   751 & 24.5201843 & $-$21.9325380 & 141.6 & 3.7 &  44.7 \\ 
  82   &   873 & 24.5136363 & $-$21.9244078 & 237.7 & 1.8 &   127.5 \\
\hline                  
\end{tabular}
\tablefoot{We identify the galaxies included in this catalogue with their photometric ID \citep[first column; see][]{ertl25} and MUSE ID (ID\textsubscript{MUSE} in the second column; see Table \ref{tab:z_muse}). We report the measured stellar velocity dispersion value of the galaxy with $\sigma_v$ (fifth column) and its uncertainty with $\delta \sigma_v$ (sixth column), and the spectral $S/N$ (seventh column). We mark the BCG with an asterisk. The horizontal line separates cluster members from background galaxies.}
\end{table*}

In Table \ref{tab:vdispcat} we present our catalogue of reliable velocity dispersion measurements, comprising 14 red cluster galaxies and two red line-of-sight background galaxies, while in Fig. \ref{fig:vdisp_spectra} we showcase the fits of their spectra performed with \texttt{pPXF}. Given the large apertures used for the spectrum extraction, we visually checked all spectra to avoid contamination from any nearby object, excluding all instances of contaminated or blended spectra, which are frequent in the most crowded regions of the cluster, besides all spectra with $S/N<10$. The spectrum of the BCG clearly shows several prominent gas emission lines (see Fig. \ref{fig:vdisp_spectra}). We find that the lowest contamination is obtained extracting the spectrum within a small circular aperture (with a $0.6''$ radius) from the BCG centre of light, avoiding the light-weighting procedure. Given that the observed emission lines influence some of the stellar absorption features used for our stellar kinematic studies, we masked out all the spectral regions within the wavelength range considered potentially affected by them, and we imposed a conservative $5\%$ uncertainty on the BCG stellar velocity dispersion value, as reported in Table \ref{tab:vdispcat}, where the BCG is marked with an asterisk. As discussed, the total uncertainty $\delta \sigma_v$ on the velocity dispersion value is obtained as the quadrature sum between the uncertainty provided by \texttt{pPXF} as a result of the spectral fit and the uncertainty associated with the $S/N$.

\subsection{The Faber-Jackson relation for the cluster members}\label{sec:fjcal}

The total mass distribution of member galaxies in cluster-scale lensing models is typically parametrised by a truncated dual pseudo-isothermal elliptical mass distribution \citep[dPIE;][]{eliasdottir07, suyu10} with vanishing core radius. Their total mass distribution thus only depends on two parameters: the velocity dispersion $\sigma_v$, and the truncation (or half-mass) radius $r_\mathrm{t}$. Cluster-scale strong lensing events produce most frequently multiple image systems with separations of the same order of magnitude as the cluster Einstein radius, typically higher than $10''$. In these cases, the positions of the multiple images are not very sensitive to the details of the total mass distribution of the members. The velocity dispersion and the truncation radius of the cluster members are thus strongly degenerate \citep{bergamini19}, as the same galaxy total mass can be obtained with a higher (lower) velocity dispersion and a lower (higher) truncation radius, leading to a more (less) compact total mass distribution for the members. To reduce the number of free parameters during the model optimisation, allowing for a positive number of degrees of freedom, the two quantities are often linked to the observed total luminosity $L$ of the members by calibrating power-law scaling relations. 

As found by \cite{faber76}, a power-law relation holds between the central stellar velocity dispersion of elliptical galaxies $\sigma_{v,0}$ and their total luminosity $L$ in a given band. As shown by \cite{bergamini19}, for a vanishing core radius, the velocity dispersion parameter of the dPIE profile is well approximated by the value of the central line-of-sight stellar velocity dispersion. As such, the calibrated Faber-Jackson relation can be used to obtain an observational prior on the corresponding scaling law used to link $\sigma_v$ and $L$. As detailed in Sect. \ref{sec:kinm}, we have accurate measurements of the central stellar velocity dispersion for 13 red cluster galaxies, excluding the BCG, which has a different morphology compared to other elliptical galaxies and is not necessarily well described by the same Faber-Jackson relation. \cite{ertl25} have measured the structural parameters of these objects in the HST band F160W. The total magnitude of the members in this band $m_{\rm F160W}$ is a good proxy of the total stellar mass of the members \citep{grillo15}. We chose the second brightest cluster member, ID\textsubscript{MUSE}=1308 with $m_\mathrm{F160W}=17.985,$ as a reference and write the velocity dispersion for the $i$-th cluster member with magnitude $m_{\mathrm{F160W},i}$ as 
\begin{equation}\label{fj}
    \sigma_{v,i}= \sigma_{v,\mathrm{ref}} \times 10^{0.4 \, (17.985-m_{\mathrm{F160W},i}) \, \alpha},
\end{equation}
where $\sigma_{v,\mathrm{ref}} $ and $\alpha$ are parameters of the scaling relation that we will determine based on our data.
\begin{figure}
        \centering
        \includegraphics[width = 1.0\columnwidth]{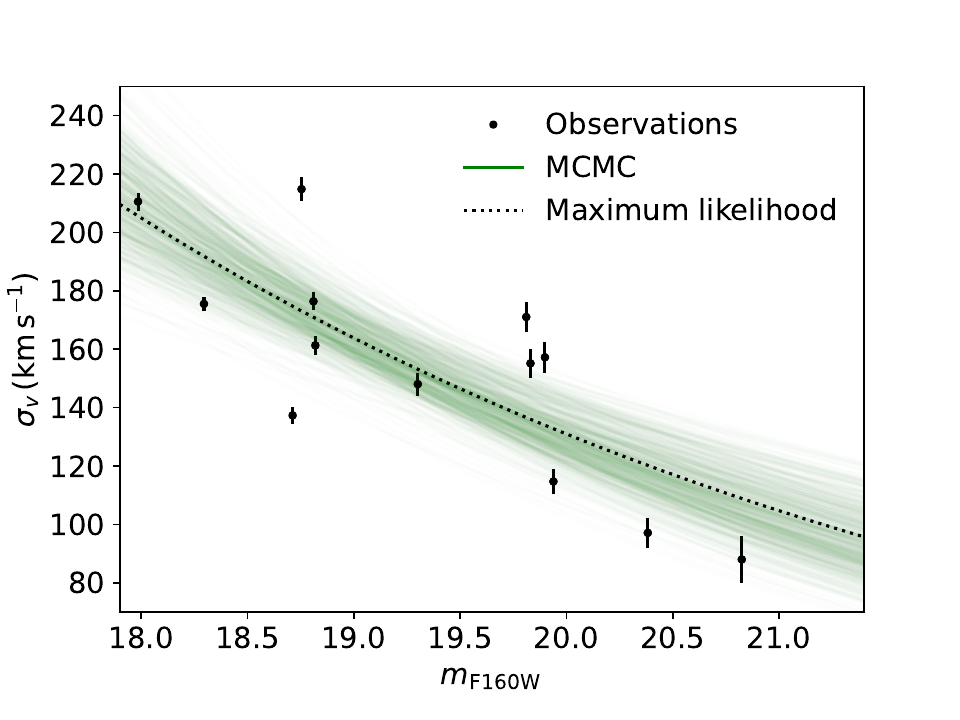}
        \caption{Faber-Jackson relation for the members of MACS0138. The green lines represent 100 realisations randomly drawn from the posterior probability distribution of the scaling relation parameters, while the dotted black line shows the maximum-likelihood solution. The black dots are the 13 cluster members with an observed velocity dispersion used for the scaling law calibration.}
        \label{fig:fj}
\end{figure}
We followed the procedure detailed in \cite{bergamini19} (Appendix B), and fit Eq. (\ref{fj}) with affine-invariant Markov chain Monte Carlo \citep[MCMC;][]{goodman10,foreman13}. We explored the posterior probability distribution for the values of $\sigma_{v,\mathrm{ref}}$ and $\alpha$ as defined in Eq. (\ref{fj}), and of the intrinsic scatter around the Faber-Jackson relation $\Delta \sigma_v$. The prior and likelihood functions, and the $\Delta \sigma_v$ are defined as in Eqs. (B.3) and (B.4) in \citet{bergamini19}. The recovered value of $\sigma_{v,\mathrm{ref}}$ is compatible with the measured $\sigma_v$ value for the reference galaxy (ID\textsubscript{MUSE}=1308; see Table \ref{tab:vdispcat}). In Fig. \ref{fig:fj} we show the Faber-Jackson relation and its uncertainty, and the 13 cluster members used to optimise its parameters. These parameters are used in \cite{ertl25} to describe the cluster-member component in their SL model of MACS0138 more accurately. In Table \ref{tab:fj} we report the median values, with a $1\sigma$ uncertainty, of the three optimised parameters, as a result of the posterior probability distribution exploration with MCMC.

\begin{table}
\caption{\label{tab:fj}Parameters of the Faber-Jackson relation for the members of MACS0138.}
\centering
\begin{tabular}{lc}
\hline\hline
Parameter&Value\\
\hline
$\sigma_{v,\mathrm{ref}} \, \mathrm{(km \, s^{-1})}$ & $206^{+14}_{-13}$ \vspace{2 pt} \\ 
$\alpha$ & $0.25^{+0.05}_{-0.05}$ \vspace{2 pt} \\
$\Delta \sigma_v  \, \mathrm{(km \, s^{-1})}$ & $25^{+6}_{-4}$ \vspace{2 pt}\\
\hline
\end{tabular}
\tablefoot{We show the median, 16th, and 84th percentile of the three optimised parameters, as defined in the main text.}
\end{table}

We compared the Faber-Jackson relation found for MACS0138 with other SL galaxy clusters within the same redshift range, as calibrated by recent SL modelling in an effort to enhance the description of the cluster member total mass component. Specifically, we considered the massive clusters Abell 2744 \citep[$z$=0.31, hereafter A2744;][]{bergamini22}, Abell S1063 \citep[$z$=0.35, hereafter AS1063;][]{bergamini19,mercurio21}, MACS J0416.1-2403 \citep[$z$=0.40, hereafter MACS0416;][]{bergamini23}, MACS J1206.2-0847 \citep[$z$=0.44, hereafter MACS1206;][]{bergamini19}, SDSS J2222+2745 \citep[$z$=0.49, hereafter SDSS2222;][]{acebron22b}, and SDSS J1029+2623 \citep[$z$=0.59, hereafter SDSS1029;][]{acebron22a}. Looking at the slope $\alpha$ of the Faber-Jackson, defined in Eq. (\ref{fj}), we notice how most values agree with our findings, with a few exceptions. The kinematic studies previously listed find $\alpha$ values of $0.40$, $0.27$, $0.30$, $0.27$, $0.295$, and $0.39$ for A2744, AS1063, MACS0416, MACS1206, SDSS2222, and SDSS1029, respectively. These values are generally slightly higher, although compatible, with what we find, except for A2744 and SDSS1029, which show a steeper slope, less consistent with the proportionality between the total galaxy luminosity and $\sigma_v^4$ found by \cite{faber76}. To perform a meaningful comparison, we used the absolute magnitudes in the Two Micron All-Sky Survey (2MASS) J band. The conversion between the rest-frame J band and the observed-frame F160W band for all seven clusters was obtained with a K-correction based on a simulated early-type galaxy spectrum generated through the GALaxy EVolution \citep[GALEV;][]{kotulla09} evolutionary synthesis models. 

\begin{figure}
        \centering
        \includegraphics[width = 1.0\columnwidth]{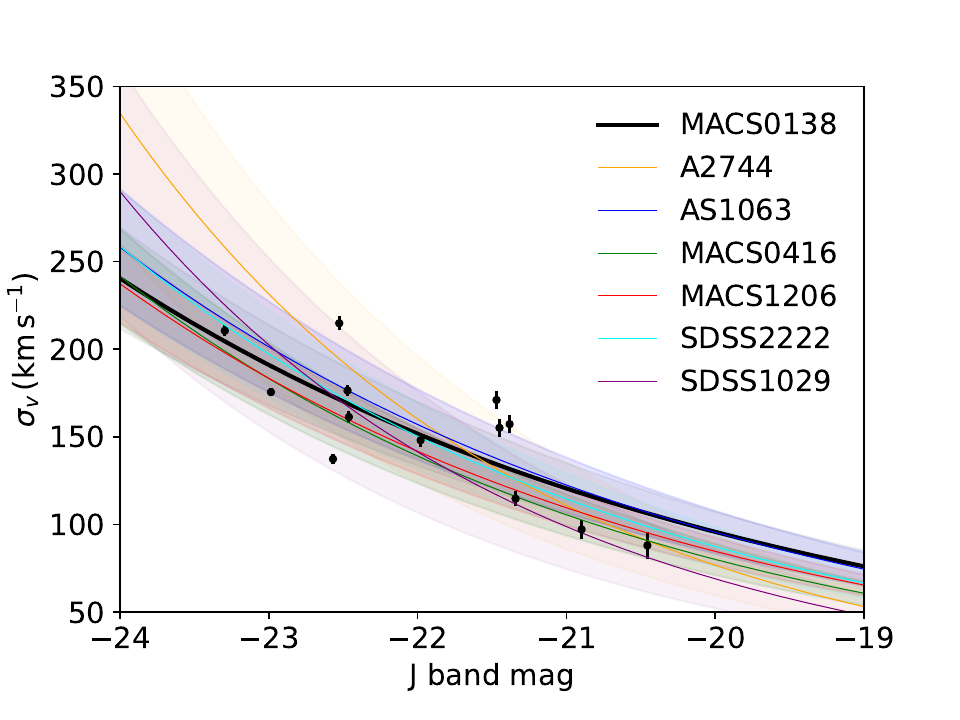}
        \caption{Faber-Jackson relation for the members of MACS0138 compared with those for six SL clusters at $0.31\leq z \leq 0.59$. We show the best-fit relation and its scatter for MACS0138 (in black), A2744 (in orange), AS1063 (in blue), MACS0416 (in green), MACS1206 (in red), SDSS2222 (in cyan), and SDSS1029 (in purple). The black dots are the 13 cluster members selected in Sect. \ref{sec:fjcal}. }
        \label{fig:fjconf}
\end{figure}

In Fig. \ref{fig:fjconf} we show the measured values of $\sigma_v$ for MACS0138 and the Faber-Jackson relations for MACS0138 and the six clusters listed above. The plot shows that all relations describe well our observed data points within their scatter, even those with a significantly higher slope, which we find to be typically associated with a higher scatter about the relation. This suggests that we do not see statistically significant signs of evolution of the Faber-Jackson relation for early-type cluster members in the redshift range $0.31\leq z \leq 0.59$, when considering total magnitude values in the same K-corrected rest-frame J band.

\section{Summary}\label{sec:conc}

We have presented a comprehensive spectroscopic analysis of the strong lensing galaxy cluster MACS0138 and an accurate study of the stellar kinematics of its bright cluster members. All our results are based on MUSE observations of the cluster core, including our recent 2.9-hour observations, for a total of approximately 3.7 hours of exposure time. In Table \ref{tab:z_muse} we present a catalogue of reliable spectroscopic redshifts ($\rm QF \ge 2$) for 107 objects. Specifically, we identify 50 galaxy cluster members at $0.324 < z < 0.349$ and a small background group at $z \sim 0.37$. Our secure identification of the mass components along the line of sight is the basis for accurate SL modelling of the cluster, as shown by \cite{ertl25}, \cite{acebron25}, and Suyu et al. (in prep.). The catalogue also contains a lensed arc-like structure at $z=1.945$ 
% \sherry{should this be 1.949 (last digit as 9) for the SN host?} 
and 13 spectroscopically confirmed ($\rm QF \ge 2$) lensed multiple images from four background sources at $0.767 \leq z \leq 3.420$. Amongst these, we report 4 lensed images of the host of SN Encore and SN Requiem, and 11 images from 
%three newly identified lensed background sources. 
three other lensed background sources, 2 of which are newly discovered here.
Our new sample of multiply lensed sources in a large redshift range provides accurate constraints on the cluster mass distribution in SL models, which will be necessary for obtaining a reliable $H_0$ measurement from the observed SN time delays.

The depth of the MUSE data has allowed us to accurately probe the stellar kinematics of several bright galaxies. In Table \ref{tab:vdispcat} we present measured stellar velocity dispersion values for 14 cluster members and 2 background galaxies, separated by a horizontal line. They have been used as a prior for the description of the galaxy-scale mass components in the SL models presented in \cite{ertl25}, \cite{acebron25}, and Suyu et al. (in prep.), breaking the parametric degeneracy that typically affects SL models of clusters on these scales. We combined our $\sigma_v$ measurements for 13 early-type cluster galaxies with their total magnitude in the HST F160W band, measured by \cite{ertl25}, to calibrate the Faber-Jackson relation for the cluster members. We find a Faber-Jackson slope $\alpha=0.25^{+0.05}_{-0.05}$, as defined in Eq. (\ref{fj}), and a scatter of $25^{+6}_{-4} \, \rm km \, s^{-1}$ about the relation. We compared the Faber-Jackson relation calibrated for MACS0138 with six other SL clusters at $0.31\leq z \leq 0.59$, finding compatible slope values (with two exceptions) and that all scaling laws seem to describe our observed data points within their scatter well. When considering total magnitude values in the same K-corrected rest-frame J band, no clear evidence of an evolution of the relation with redshift is observed.

During the referee process for this paper, \cite{flowers24} published in a pre-print an independent spectroscopic redshift catalogue of MACS0138 and the calibration of the Faber-Jackson relation for its members. Their work differs from ours as they used shallower MUSE data (i.e. 49 minutes from the ESO programme ID 0103.A-0777), leading to a smaller sample of 30 galaxies in the field with redshift measurements, including 23 cluster members, all of which are included in this work. Despite the shallower data, their velocity dispersion catalogue is instead slightly larger, comprising 19 objects, likely due to different selection cuts. Whilst we simultaneously fit for the stellar population and the velocity dispersion of each galaxy, the velocity dispersion measurements in \cite{flowers24} were performed with a single spectral template, built from a galaxy chosen amongst the brightest in the cluster. \cite{flowers24} tested the impact of this and other modelling configuration choices on the velocity dispersion results, confirming its robustness. The Faber-Jackson relation is calibrated with respect to the magnitude measured in a bluer band (F555W instead of F160W), but the slope, normalisation, and scatter are consistent with our results, supporting the robustness of the kinematic calibration of the Faber-Jackson relation. However, a few values reported in \citet{flowers24} differ very significantly from our measurements, such as the cluster member $\mathrm{ID_{MUSE}}=930$ and the background galaxy $\mathrm{ID_{MUSE}}=873$, for which they report velocity dispersion values of $167 \pm 15 \, \rm km \, s^{-1}$ and $291 \pm 3 \, \rm km \, s^{-1}$, whilst we find $97 \pm 5 \, \rm km \, s^{-1}$ and $238 \pm 2 \, \rm km \, s^{-1}$, respectively. A possible explanation for their significantly higher $\sigma_v$ values is that their noisier spectra, owing to shallower observations, hinder the full recovery of the absorption line profile, as suggested by Fig. 7 in \cite{flowers24}.

In conclusion, the new MUSE observations have paved the way for a deep analysis of the redshift-space distribution of cluster members and line-of-sight objects in the field of the cluster MACS0138, including the identification of multiply lensed images of emission-line sources within a large redshift range. We have furthermore shown the importance of MUSE observations for the study of the stellar kinematics of a sizeable sample of cluster and background galaxies. Our analysis will be key for the development of a set of accurate SL models of MACS0138, which is necessary to obtain a measurement of $H_0$ from the time delays between the multiple images of the lensed SN. The procedure and techniques presented in this work will prove extremely valuable in the upcoming years, as the Legacy Survey of Space and Time at the \textit{Vera C. Rubin} Observatory will discover thousands of time-variable sources lensed by galaxies and clusters of galaxies \citep{oguri10}.

\begin{acknowledgements}
We thank Tommaso Treu and Shawn Knabel for the help and useful discussions on the velocity dispersion measurements. We thank Josè M. Diego and Conor Larison for their useful comments. GG and CG acknowledge support from the Italian Ministry of University and Research through grant PRIN-MIUR 2020SKSTHZ. SE and SHS thank the Max Planck Society for support through the Max Planck Fellowship for SHS. This work is supported in part by the Deutsche Forschungsgemeinschaft (DFG, German Research Foundation) under Germany's Excellence Strategy -- EXC-2094 -- 390783311. SS has received funding from the European Union’s Horizon 2022 research and innovation programme under the Marie Skłodowska-Curie grant agreement No 101105167 — FASTIDIoUS. 
This project has received funding from the European Research Council (ERC) under the European Union’s Horizon 2020 research and innovation programme (grant agreement No 771776). 
This work uses the following software packages:
\href{https://github.com/astropy/astropy}{\texttt{Astropy}}
\citep{astropy1, astropy2},
\href{https://github.com/matplotlib/matplotlib}{\texttt{matplotlib}}
\citep{matplotlib},
\href{https://github.com/numpy/numpy}{\texttt{NumPy}}
\citep{numpy1, numpy2},
\href{https://pypi.org/project/ppxf/}{\texttt{pPXF}}
\citep{cappellari04,cappellari23},
\href{https://www.python.org/}{\texttt{Python}}
\citep{python},
\href{https://github.com/scipy/scipy}{\texttt{Scipy}}
\citep{scipy}.

\end{acknowledgements}

\bibliographystyle{aa}
\bibliography{references}

\clearpage

\begin{appendix}
\onecolumn
\section{MUSE spectroscopic redshift catalogue}

In this appendix we present the final MUSE spectroscopic redshift catalogue, containing 107 reliable (i.e. likely or secure $\rm QF \ge 2$) redshift measurements of extragalactic objects, with two sources in the foreground, 58 at the cluster redshift, including 50 cluster galaxies, and 47 in the cluster background, as discussed in Sect. \ref{sec:zcat}.

\begin{longtable}{cccccc}
\caption{\label{tab:z_muse} MUSE spectroscopic redshift catalogue.}\\
\hline\hline
ID & ID\textsubscript{MUSE}& R.A. & Dec & $z$ & QF \\  & & (J2000) & (J2000) &  & \\
\hline
\endfirsthead
\caption{Continued.}\\
\hline\hline
ID & ID\textsubscript{MUSE}& R.A. & Dec & $z$ & QF \\  & & (J2000) & (J2000) &  & \\
\hline
\endhead
\hline
\endfoot
-- & 9900008 & 24.5229109 & $-$21.9260049 & 0.2789 & 9 \\ 
-- & 555552 & 24.5159631 & $-$21.9300671 & 0.3092 & 3 \\ \hline
117 & 1738 & 24.5204485 & $-$21.9197347 & 0.3285 & 3 \\ 
-- & 1716$^*$ & 24.5232187 & $-$21.9209395 & 0.3291 & 3 \\ 
320 & 1162 & 24.5178425 & $-$21.9303273 & 0.3307 & 2 \\ 
-- & 10002 & 24.5161968 & $-$21.9247577 & 0.3307 & 3 \\ 
-- & 10001 & 24.5172345 & $-$21.9235510 & 0.3311 & 3 \\ 
78 & 1127 & 24.5149459 & $-$21.9276509 & 0.3316 & 3 \\ 
-- & 1619$^*$ & 24.5175558 & $-$21.9223099 & 0.3317 & 3 \\ 
-- & 555559$^*$ & 24.5171543 & $-$21.9208968 & 0.3318 & 3 \\ 
-- & 1686$^*$ & 24.5148629 & $-$21.9213703 & 0.3318 & 3 \\ 
103 & 1073 & 24.512084 & $-$21.9252872 & 0.3323 & 3 \\ 
-- & 1644$^*$ & 24.516643 & $-$21.9222194 & 0.3324 & 3 \\ 
261 & 5555521 & 24.5097565 & $-$21.9302674 & 0.3328 & 3 \\ 
154 & 1498 & 24.5129692 & $-$21.9203716 & 0.3329 & 3 \\ 
262 & 1218 & 24.5130061 & $-$21.9278434 & 0.3331 & 3 \\ 
92 & 1304 & 24.5123572 & $-$21.9253942 & 0.3332 & 3 \\ 
199 & 1782 & 24.5099978 & $-$21.9190911 & 0.3335 & 3 \\ 
60 & 997 & 24.5153411 & $-$21.9329004 & 0.3336 & 2 \\ 
115 & 1308 & 24.5114284 & $-$21.9213096 & 0.3337 & 3 \\ 
76 & 1149 & 24.5228962 & $-$21.9298747 & 0.3338 & 3 \\ 
-- & 774 & 24.5099894 & $-$21.9291747 & 0.3339 & 3 \\ 
142 & 1551 & 24.5143184 & $-$21.9224156 & 0.3341 & 3 \\ 
-- & 1316$^*$ & 24.5090635 & $-$21.9274273 & 0.3345 & 3 \\ 
-- & 1318$^*$ & 24.5089117 & $-$21.9274546 & 0.3348 & 3 \\ 
70 & 1094 & 24.5119004 & $-$21.930871 & 0.3352 & 3 \\ 
55 & 895 & 24.5162254 & $-$21.9326011 & 0.3353 & 3 \\ 
85 & 1256 & 24.510652 & $-$21.9279027 & 0.3357 & 3 \\ 
113 & 1475 & 24.5150361 & $-$21.9242069 & 0.3360 & 2 \\ 
135 & 1535 & 24.5133605 & $-$21.9225114 & 0.3360 & 3 \\ 
485 & 2000 & 24.51663 & $-$21.9244886 & 0.3363 & 3 \\ 
129 & 1513 & 24.5201735 & $-$21.9229667 & 0.3365 & 3 \\ 
93 & 555553 & 24.5141805 & $-$21.927675 & 0.3368 & 2 \\ 
267 & 1642 & 24.5128585 & $-$21.9217549 & 0.3370 & 2 \\ 
116 & 1457 & 24.5156388 & $-$21.9227696 & 0.3372 & 3 \\ 
90 & 1211 & 24.5076178 & $-$21.9271544 & 0.3372 & 3 \\ 
105 & 555554 & 24.5108496 & $-$21.9255751 & 0.3374 & 3 \\ 
102 & 1210 & 24.5213768 & $-$21.9239768 & 0.3379 & 3 \\ 
153 & 1630 & 24.5241351 & $-$21.919833 & 0.3379 & 3 \\ 
106 & 5555536 & 24.5119913 & $-$21.9254078 & 0.3380 & 3 \\ 
0 & 827 & 24.5157373 & $-$21.9254784 & 0.3381 & 3 \\ 
58 & 930 & 24.5078679 & $-$21.9329074 & 0.3383 & 3 \\ 
63 & 1049 & 24.5207142 & $-$21.9319295 & 0.3383 & 3 \\ 
72 & 1147 & 24.5113797 & $-$21.9300727 & 0.3385 & 3 \\ 
193 & 1631 & 24.5153499 & $-$21.9192529 & 0.3386 & 3 \\ 
191 & 1386 & 24.5214401 & $-$21.9183725 & 0.3386 & 3 \\ 
265 & 1783 & 24.5222923 & $-$21.9191665 & 0.3387 & 2 \\ 
77 & 555551 & 24.5177768 & $-$21.9292971 & 0.3390 & 3 \\ 
69 & 1032 & 24.5090052 & $-$21.9312181 & 0.3392 & 3 \\ 
264 & 1582 & 24.5244402 & $-$21.9220557 & 0.3394 & 2 \\ 
156 & 1633 & 24.5162417 & $-$21.9200615 & 0.3399 & 3 \\ 
65 & 1003 & 24.5059888 & $-$21.9311714 & 0.3406 & 3 \\ 
52 & 898 & 24.5257606 & $-$21.932203 & 0.3416 & 3 \\ 
192 & 1701 & 24.5186628 & $-$21.9198871 & 0.3427 & 3 \\ 
202 & 1549 & 24.5233304 & $-$21.9219431 & 0.3431 & 3 \\ 
203 & 1433 & 24.516425 & $-$21.9212544 & 0.3436 & 3 \\ 
-- & 1431$^*$ & 24.5084096 & $-$21.9253918 & 0.3437 & 3 \\ 
263 & 1543 & 24.5188906 & $-$21.9228596 & 0.3453 & 3 \\ 
266 & 1595 & 24.515279 & $-$21.9220368 & 0.3463 & 3 \\ 
189 & 1869 & 24.5140145 & $-$21.917715 & 0.3472 & 3 \\ \hline
-- & 1952 & 24.5166145 & $-$21.9171875 & 0.354 & 9 \\ 
67 & 1066 & 24.5171971 & $-$21.9320865 & 0.370 & 3 \\ 
82 & 873 & 24.5136363 & $-$21.9244078 & 0.371 & 3 \\ 
51 & 751 & 24.5201843 & $-$21.932538 & 0.372 & 3 \\ 
-- & 1070 & 24.5127061 & $-$21.9320308 & 0.372 & 3 \\ 
194 & 1608 & 24.5244187 & $-$21.9196872 & 0.376 & 3 \\ 
-- & 5555516 & 24.5132654 & $-$21.9305945 & 0.459 & 3 \\ 
-- & 9900014 & 24.5246215 & $-$21.9324916 & 0.522 & 3 \\ 
-- & 1838 & 24.5099019 & $-$21.9183388 & 0.541 & 3 \\ 
-- & 1911 & 24.5084136 & $-$21.9176089 & 0.602 & 3 \\ 
-- & 9900011 & 24.5103473 & $-$21.9294554 & 0.652 & 3 \\ 
-- & \textbf{1532} & 24.5150086 & $-$21.923182 & 0.767 & 3 \\ 
-- & \textbf{9900015} & 24.5146393 & $-$21.9262452 & 0.767 & 3 \\ 
-- & \textbf{1299}& 24.518391 & $-$21.9264267 & 0.767 & 3 \\ 
187 & 1674 & 24.5103495 & $-$21.9164811 & 0.791 & 3 \\ 
-- & 1826 & 24.5088558 & $-$21.9189801 & 0.793 & 3 \\ 
-- & 1897 & 24.5075894 & $-$21.9178839 & 0.830 & 3 \\ 
84 & 1200 & 24.5076133 & $-$21.9264195 & 0.830 & 3 \\ 
-- & 1436 & 24.50895 & $-$21.9250223 & 0.862 & 3 \\ 
-- & 1915 & 24.5085166 & $-$21.9174518 & 0.870 & 3 \\ 
-- & 1863 & 24.5163104 & $-$21.9182446 & 0.951 & 3 \\ 
183 & 1917 & 24.5205768 & $-$21.9161932 & 0.952 & 3 \\ 
-- & 1117 & 24.5133754 & $-$21.931416 & 1.024 & 3 \\ 
319 & 1093 & 24.5128482 & $-$21.931183 & 1.024 & 3 \\ 
-- & 1026 & 24.5189773 & $-$21.9332474 & 1.347 & 3 \\ 
-- & 1005 & 24.5188692 & $-$21.9332751 & 1.347 & 3 \\ 
205 & 1683 & 24.508369 & $-$21.9205608 & 1.666 & 3 \\ 
97 & 1332 & 24.5228653 & $-$21.9262572 & 1.759 & 3 \\ 
-- & 1755 & 24.5153988 & $-$21.9187909 & 1.945 & 2 \\ 
-- & \textbf{555558} & 24.5176138 & $-$21.9233427 & 1.949 & 3 \\ 
-- & \textbf{927} & 24.5132561 & $-$21.9299314 & 1.949 & 3 \\ 
-- & \textbf{364} & 24.5160096 & $-$21.9303706 & 1.949 & 3 \\ 
-- & \textbf{761} & 24.5099201 & $-$21.9260449 & 1.949 & 3 \\ 
-- & 1366 & 24.5255739 & $-$21.9265684 & 2.950 & 2 \\ 
-- & \textbf{9900003} & 24.5168229 & $-$21.9256552 & 3.152 & 3 \\ 
-- & \textbf{9900002} & 24.5180612 & $-$21.9255557 & 3.152 & 3 \\ 
-- & \textbf{9900004} & 24.5082954 & $-$21.9228268 & 3.152 & 3 \\ 
-- & 9900001 & 24.5209073 & $-$21.9318334 & 3.361 & 9 \\ 
-- & \textbf{9900005} & 24.507886 & $-$21.9242017 & 3.420 & 3 \\ 
-- & \textbf{9900006} & 24.5174131 & $-$21.9250015 & 3.420 & 3 \\ 
-- & \textbf{9900007} & 24.5158357 & $-$21.9247664 & 3.420 & 2 \\ 
-- & 9900009 & 24.5241269 & $-$21.9286028 & 3.584 & 9 \\ 
-- & 9900010 & 24.5226685 & $-$21.9224566 & 3.620 & 3 \\ 
-- & 9900012 & 24.5118702 & $-$21.9177915 & 4.068 & 3 \\ 
-- & 1175 & 24.5237221 & $-$21.9302823 & 4.354 & 2 \\ 
-- & 9900013 & 24.5238134 & $-$21.9247531 & 4.641 & 3 \\
-- & 1947 & 24.5175665 & -21.9173905 & 6.353 & 3 \\

\end{longtable}
\tablefoot{We identify the sources included in this catalogue with their photometric ID \citep[see][]{ertl25} and ID\textsubscript{MUSE}. We report the right ascension, declination, spectroscopic redshift $z$, and the corresponding uncertainty in terms of a QF, with 3 as secure, 2 as likely, and 9 as secure (based on a single emission line). The horizontal line separates objects at the cluster redshift from line-of-sight foreground and background galaxies. We mark the 13 lensed multiple images with a reliable redshift measurements (i.e. QF$\ge 2$) included in this catalogue with a bold-face ID, and the objects belonging to the cluster which are not galaxies with an asterisk (such non-galaxy objects are e.g. star-forming clumps of jellyfish galaxies at the cluster redshift). They have been excluded from the cluster member sample used by \cite{ertl25}, \cite{acebron25}, and Suyu et al. (in prep.) to build SL models of the cluster.}

\section{Measuring velocity dispersions from MUSE spectra: Spectral template fitting results}

In this appendix we show the results of the pPXF-based line-of-sight stellar velocity dispersion fitting for 14 cluster members of MACS0138 and two background galaxies from their MUSE spectra. The procedure for these measurement is detailed in Sect. \ref{sec:kin} and the results are presented in Table \ref{tab:vdispcat}. 

\FloatBarrier

\begin{figure*}[h!]
  \includegraphics[width = 0.5\columnwidth]{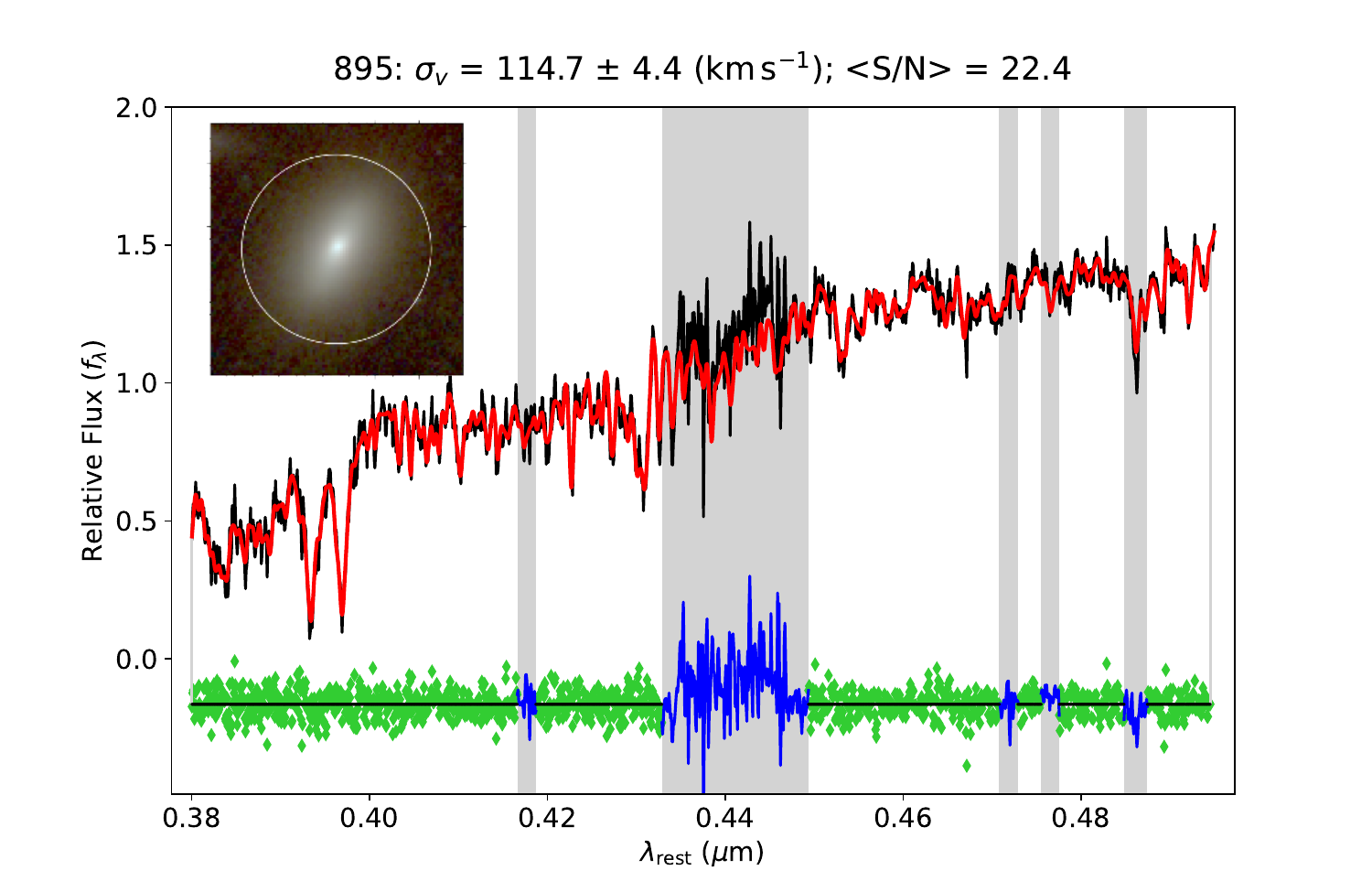}
  \includegraphics[width = 0.5\columnwidth]{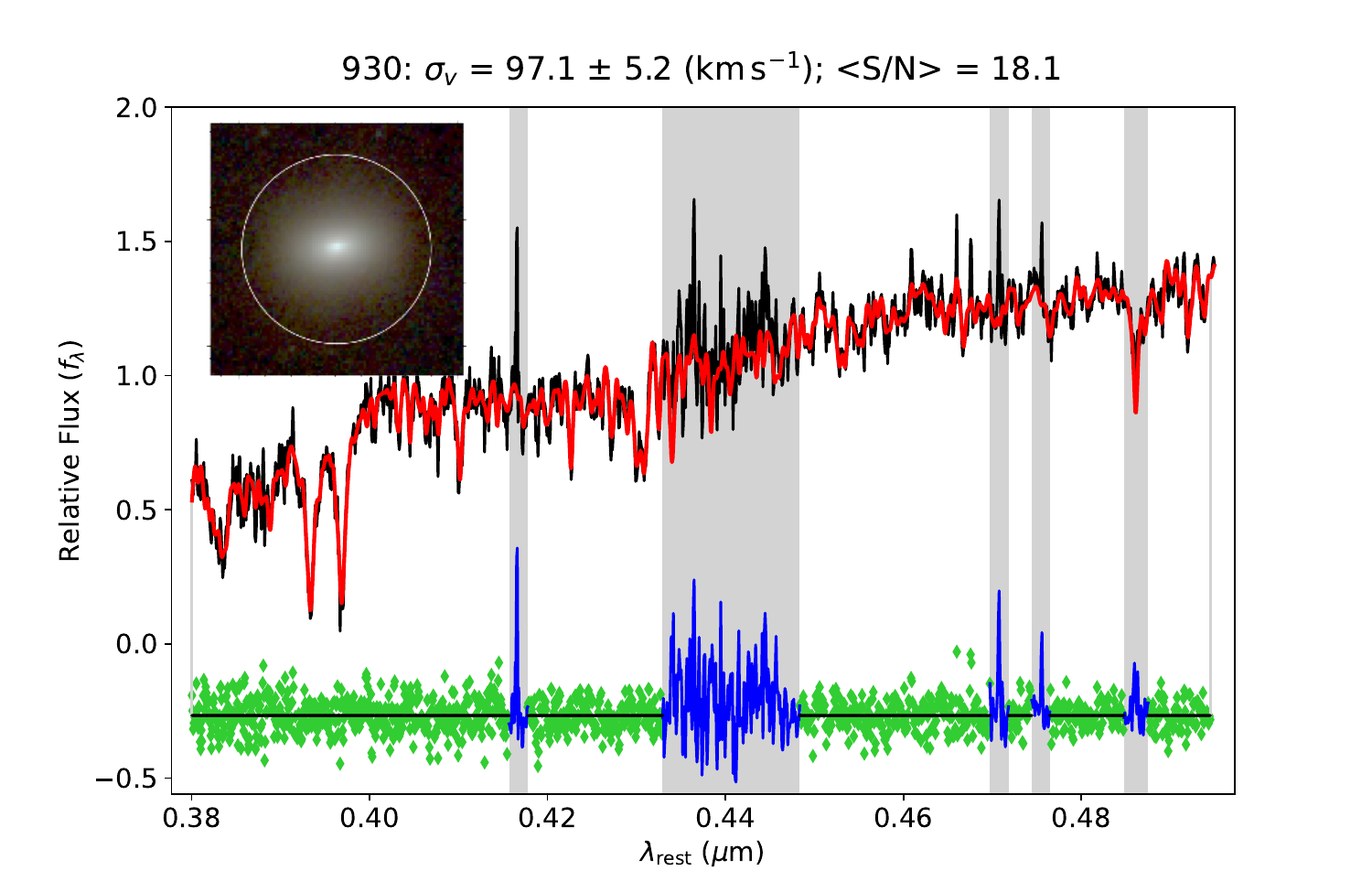}
  \includegraphics[width = 0.5\columnwidth]{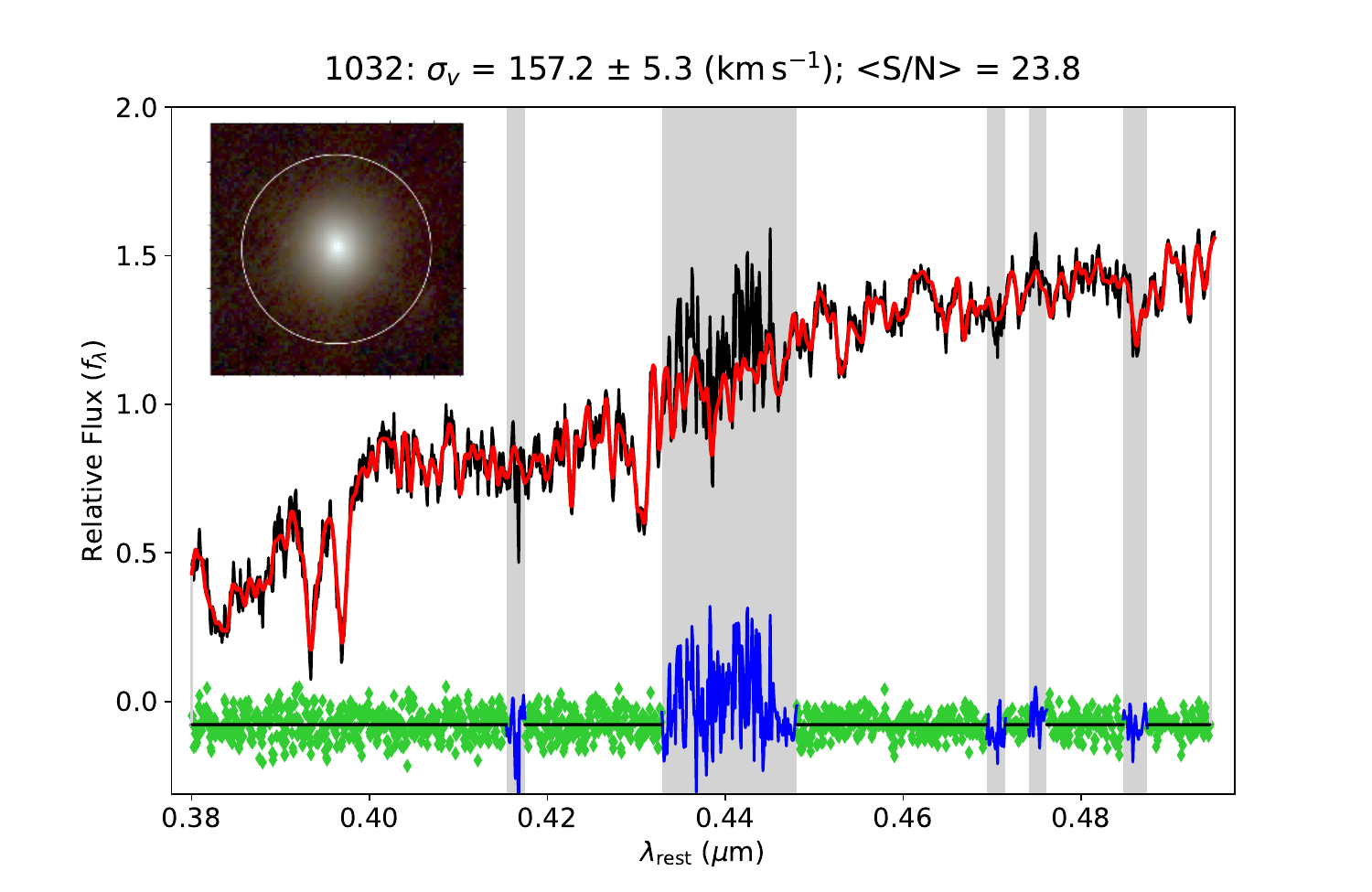}
  \includegraphics[width = 0.5\columnwidth]{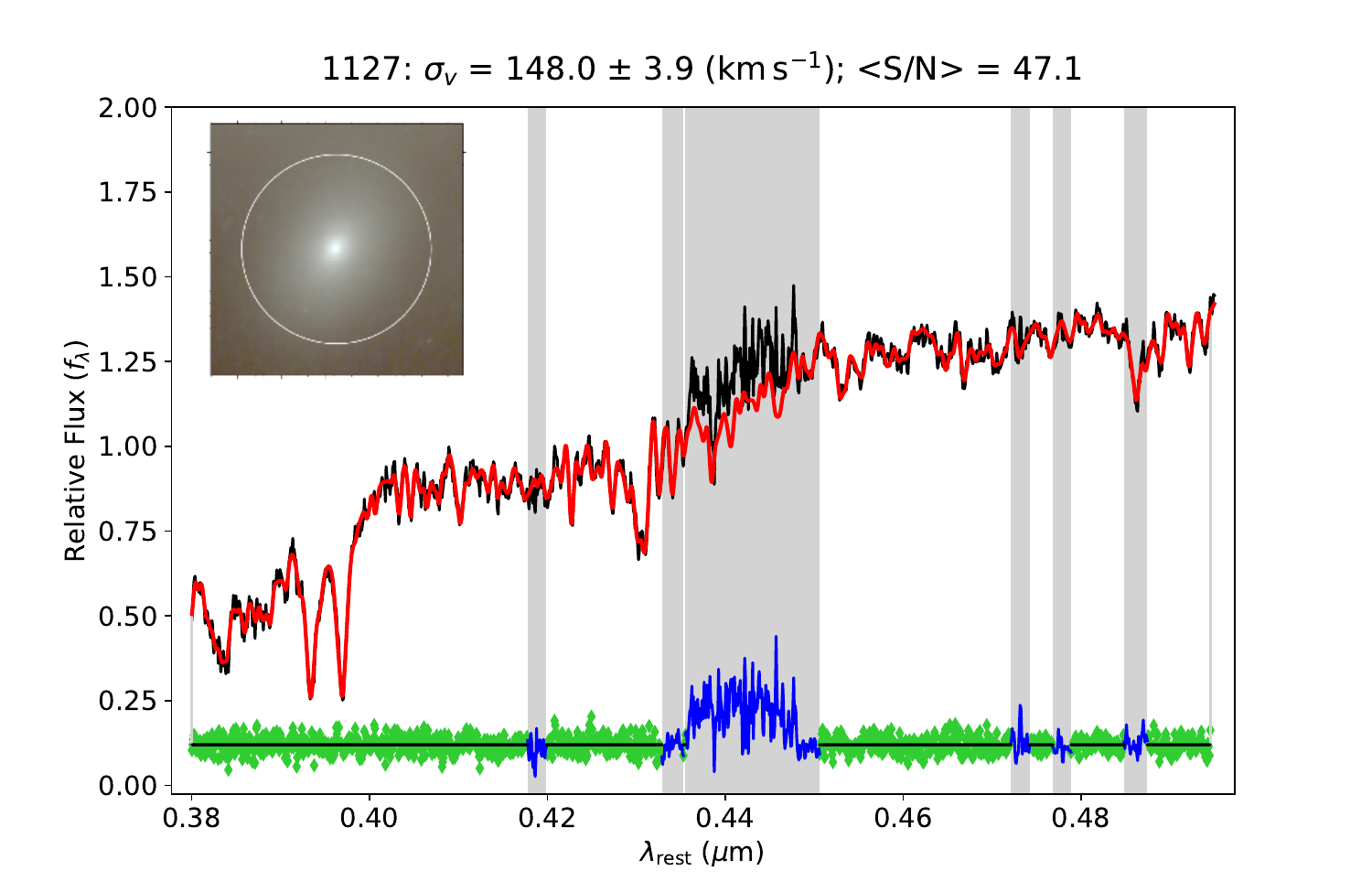}
  \caption{Stellar velocity dispersion fitting with \texttt{pPXF}. Observed MUSE spectra are in black, and the red curves are the \texttt{pPXF} best-fit models. The residuals of the data with respect to the model are shown in green. The regions masked, to avoid the impact of emission lines or of the subtraction of sky, laser, or telluric lines, are shaded in grey. The data minus model residuals in these regions are marked in blue.
  In the top-left corner, we show a JWST red-green-blue cutout of the object, $4''$ on each side, and show the apertures used for the spectral extraction from the weighted MUSE cube. Measured $\sigma_v$ and $S/N$ values are shown above the plot.}
  \label{fig:vdisp_spectra}
\end{figure*}
\begin{figure*}[h!]
  \includegraphics[width = 0.5\columnwidth]{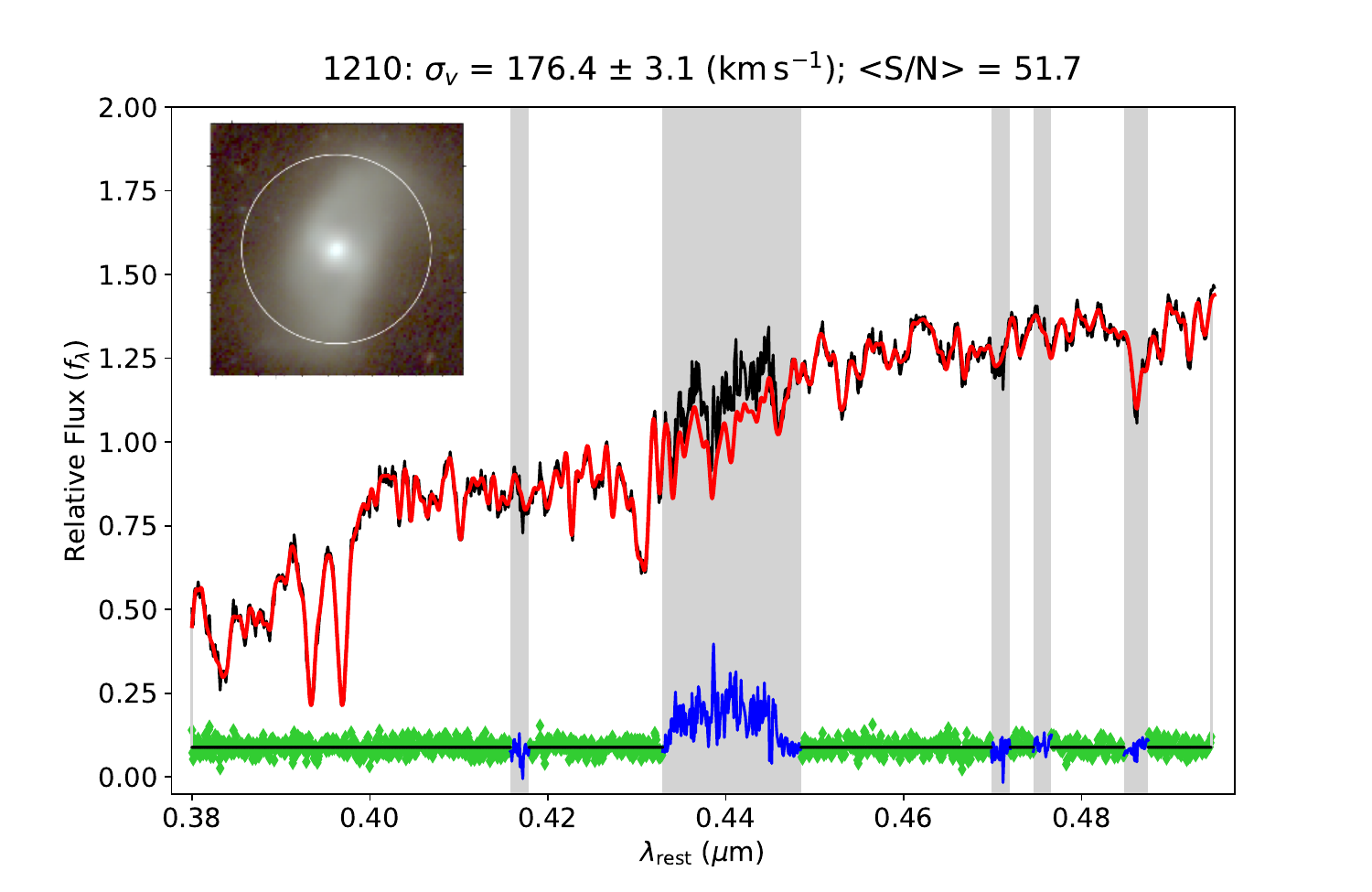}
  \includegraphics[width = 0.5\columnwidth]{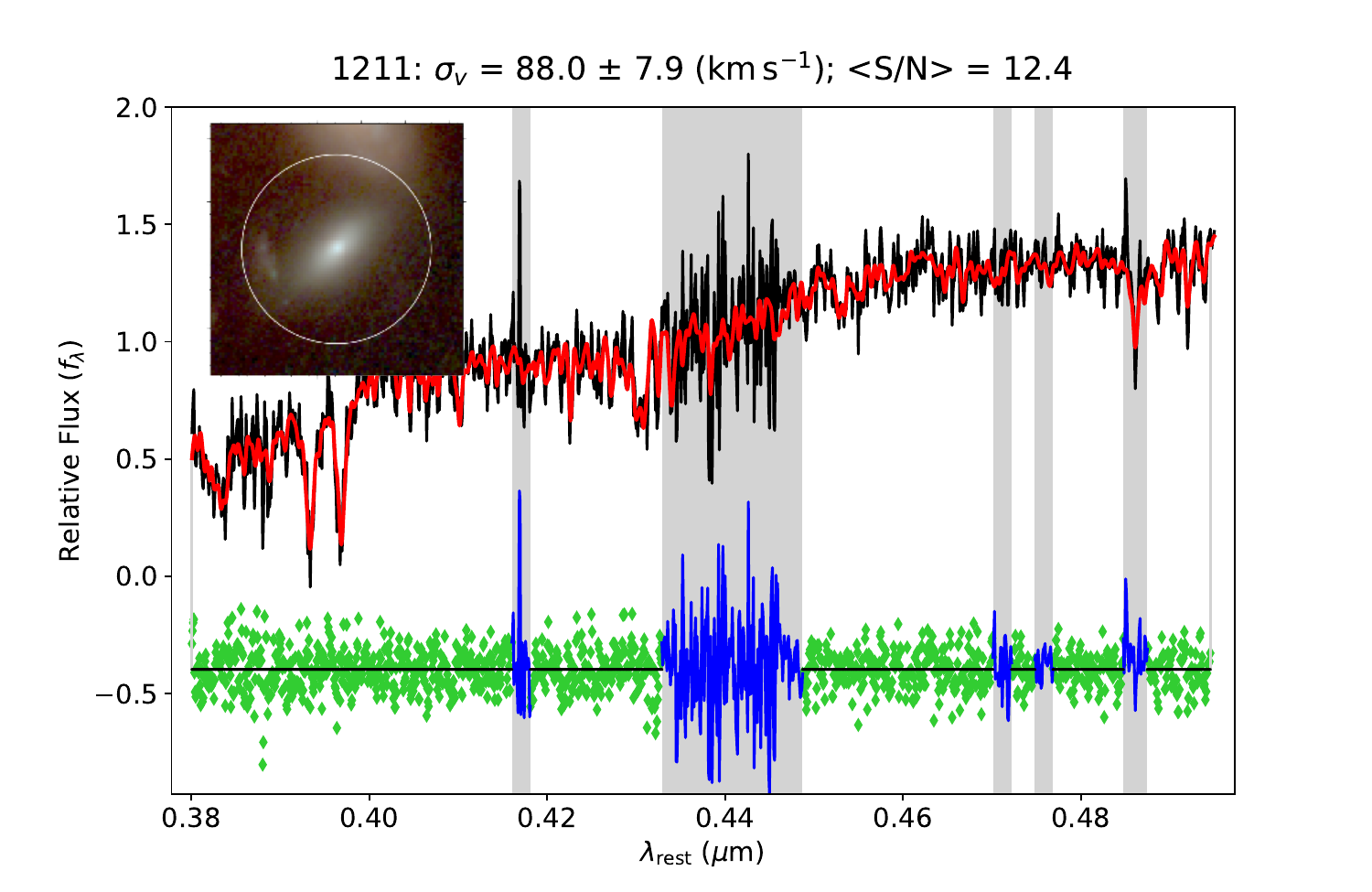}
  \includegraphics[width = 0.5\columnwidth]{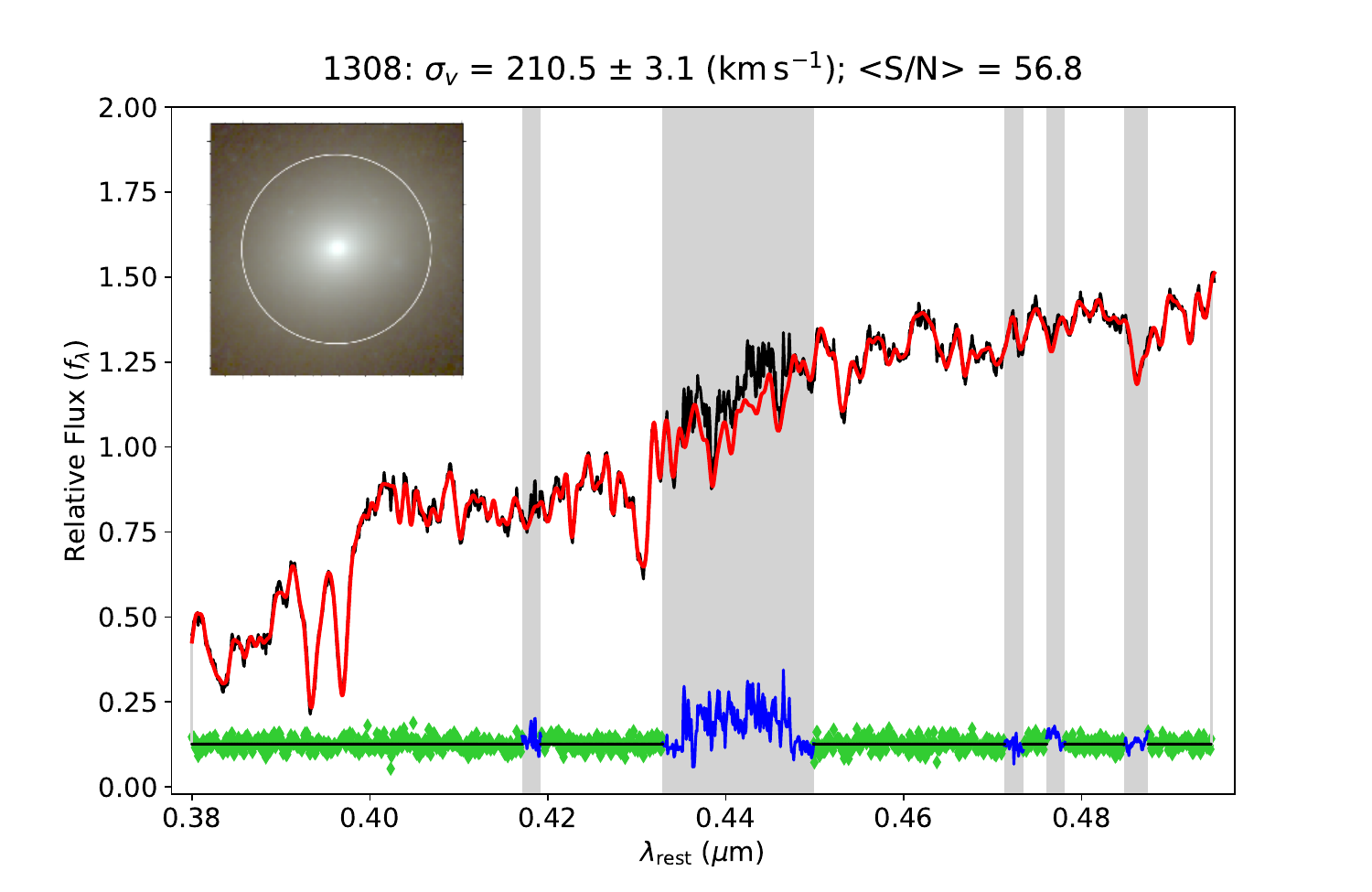}
  \includegraphics[width = 0.5\columnwidth]{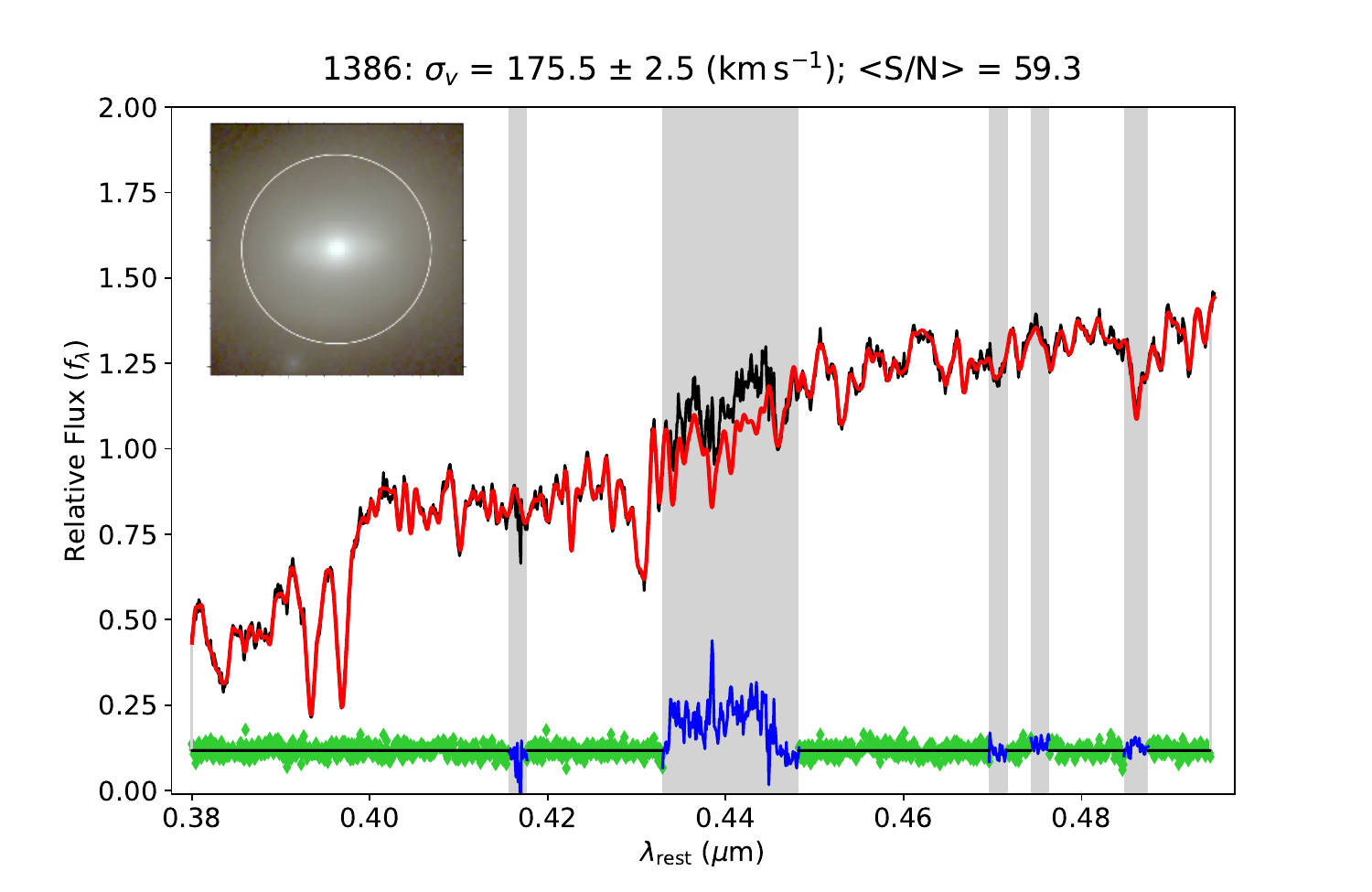}
  \includegraphics[width = 0.5\columnwidth]{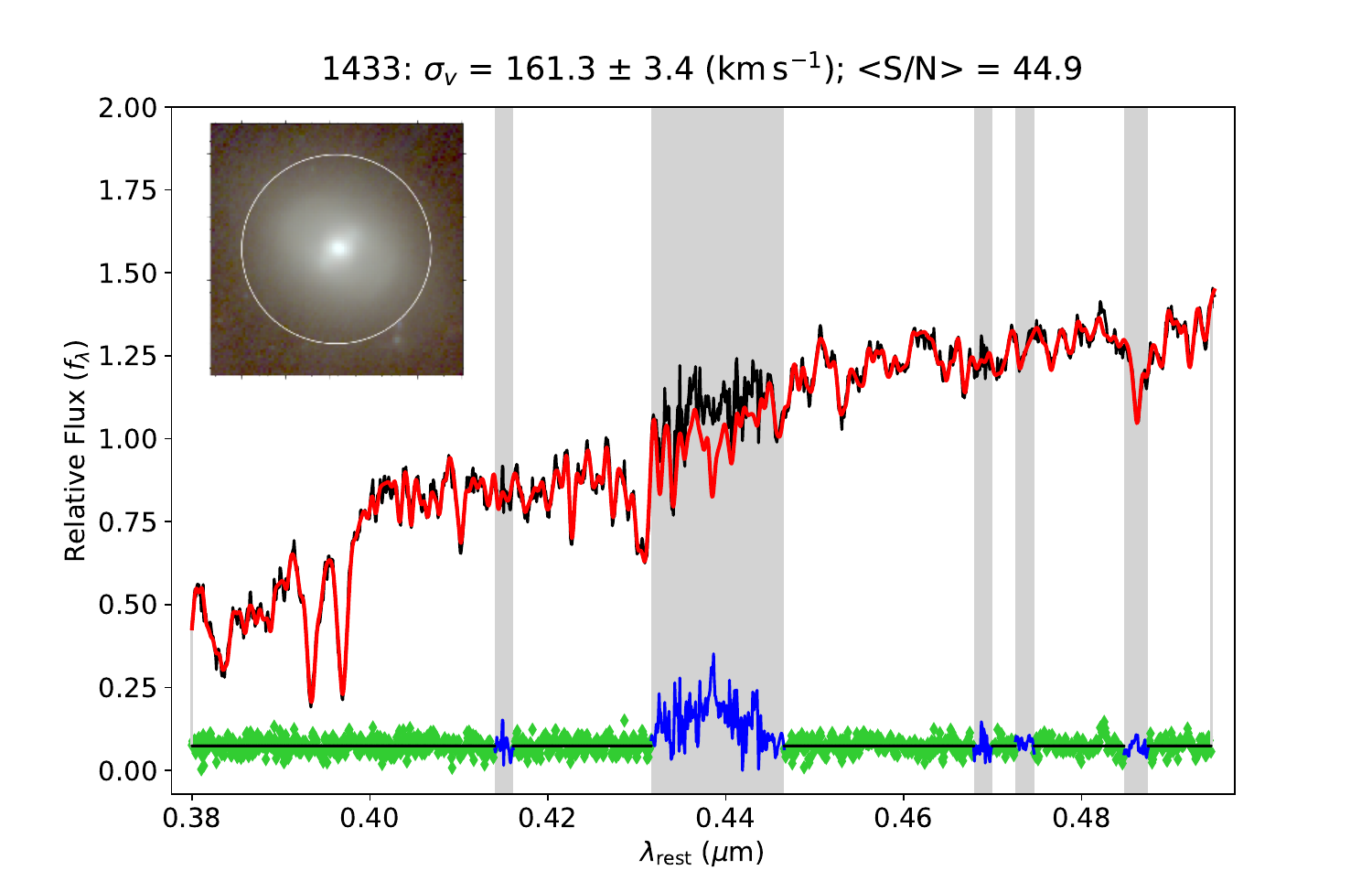}
  \includegraphics[width = 0.5\columnwidth]{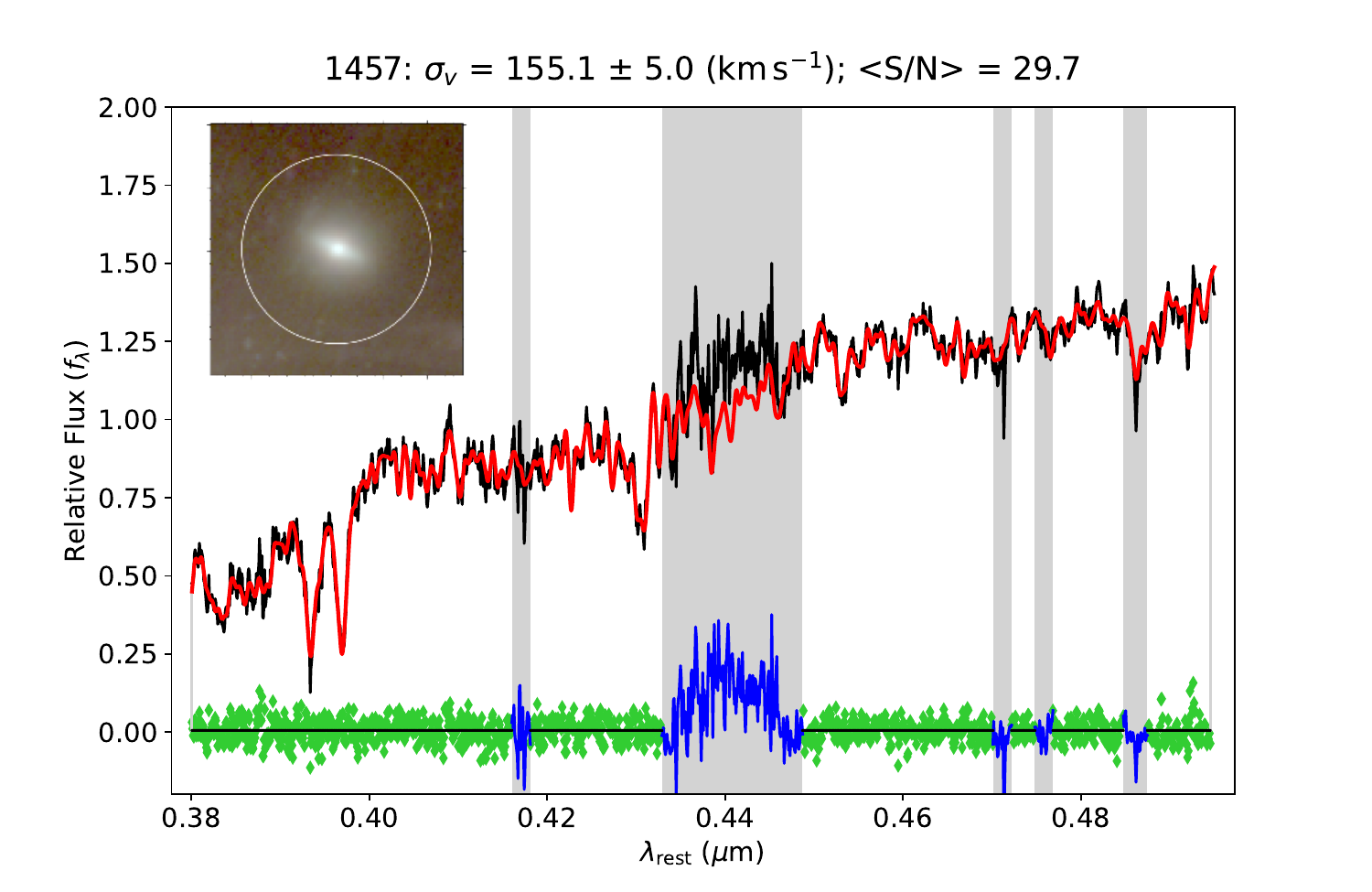}
  \caption*{Continued.}
  \label{fig:vdisp_spectra2}
\end{figure*}
\begin{figure*}[h!]
  \includegraphics[width = 0.5\columnwidth]{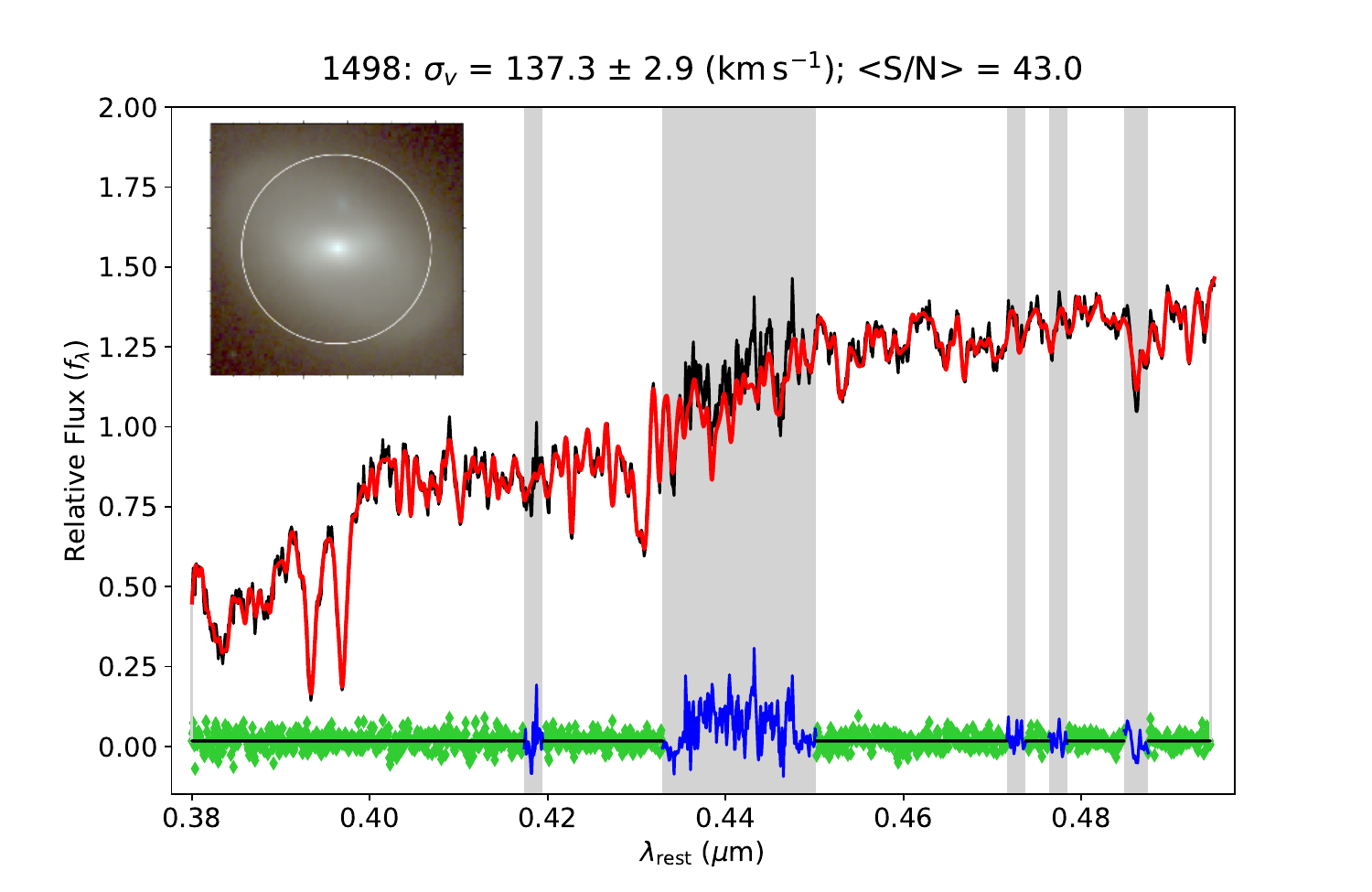}
  \includegraphics[width = 0.5\columnwidth]{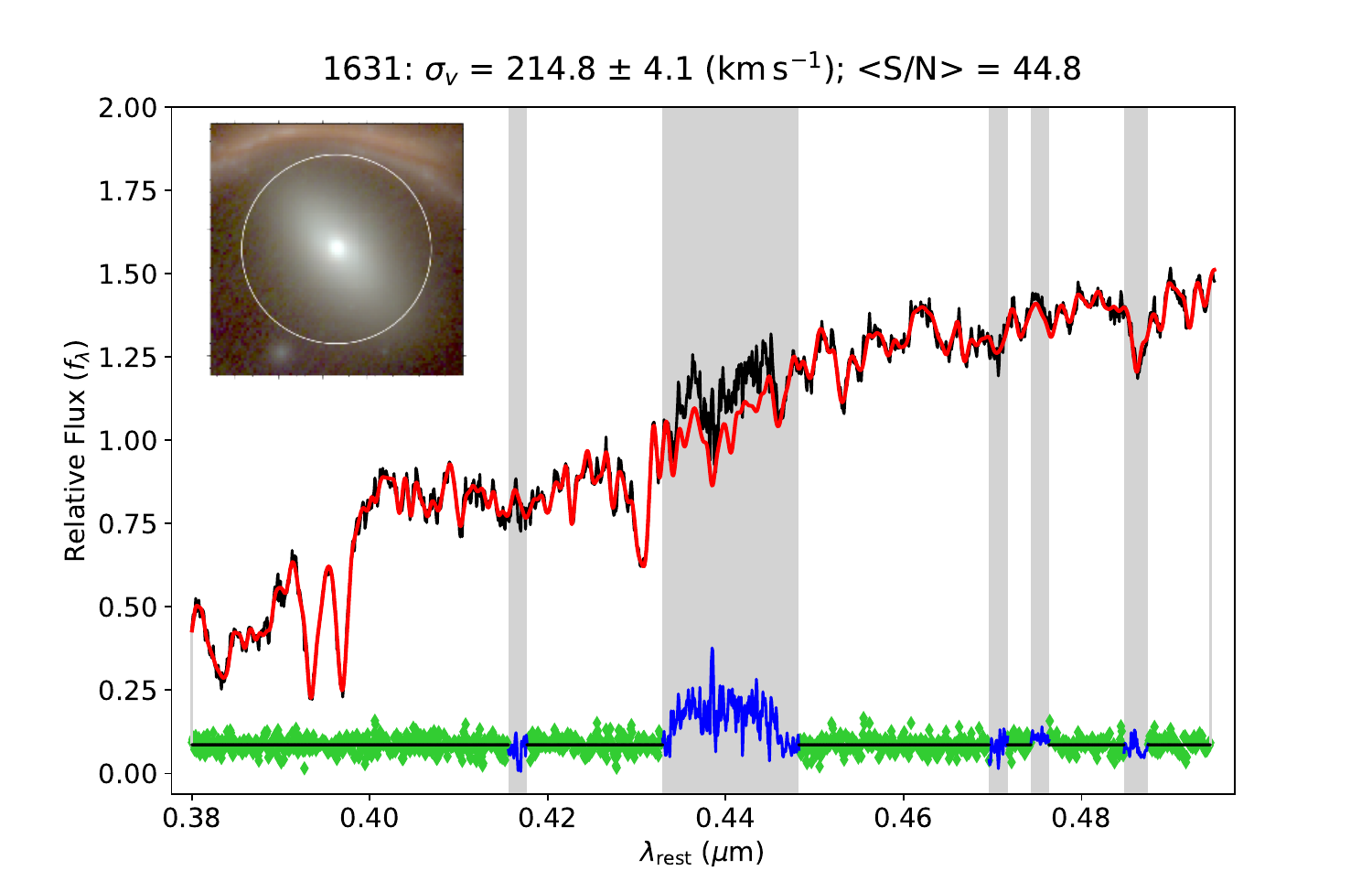}
  \includegraphics[width = 0.5\columnwidth]{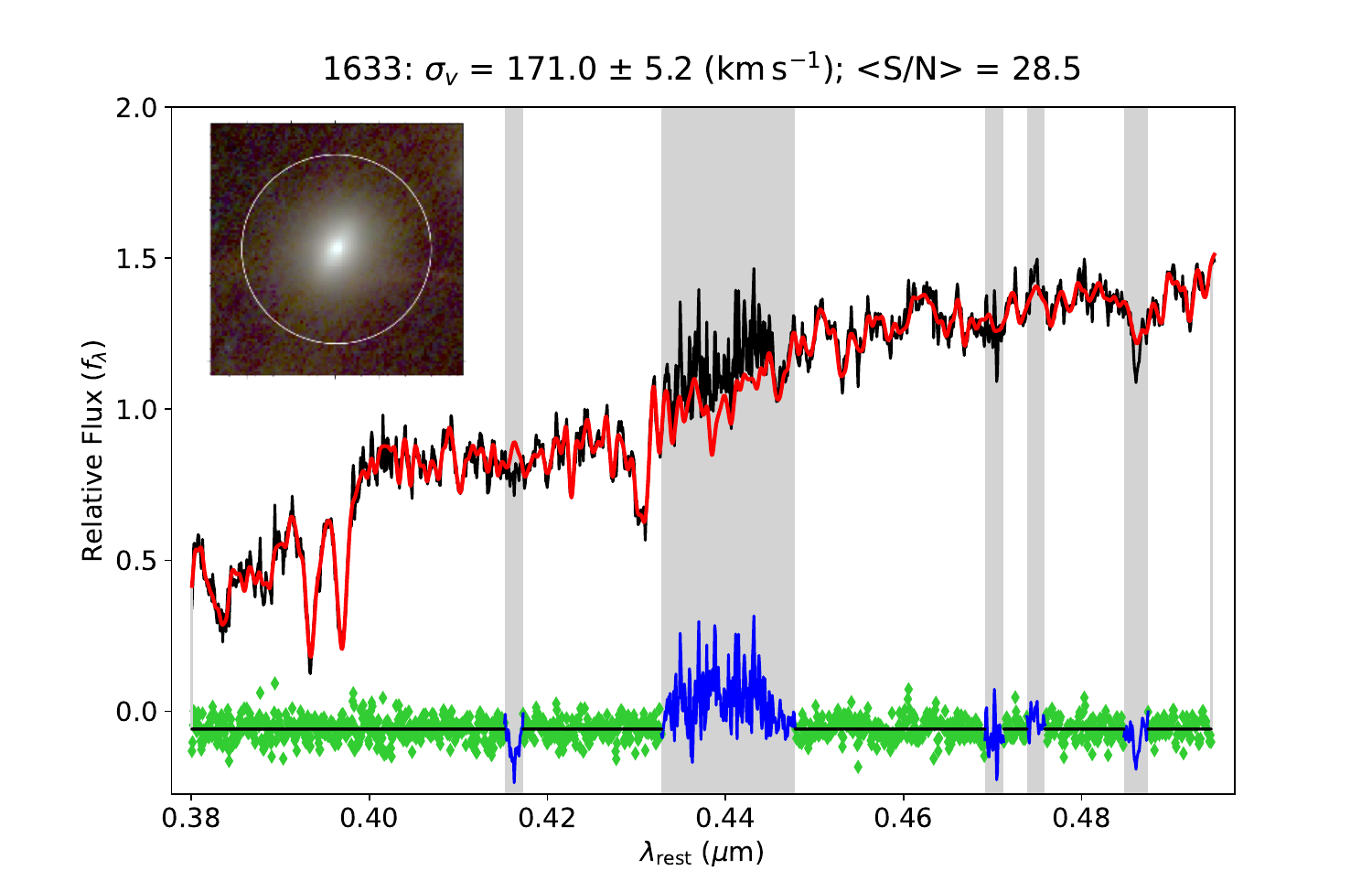}
  \includegraphics[width = 0.5\columnwidth]{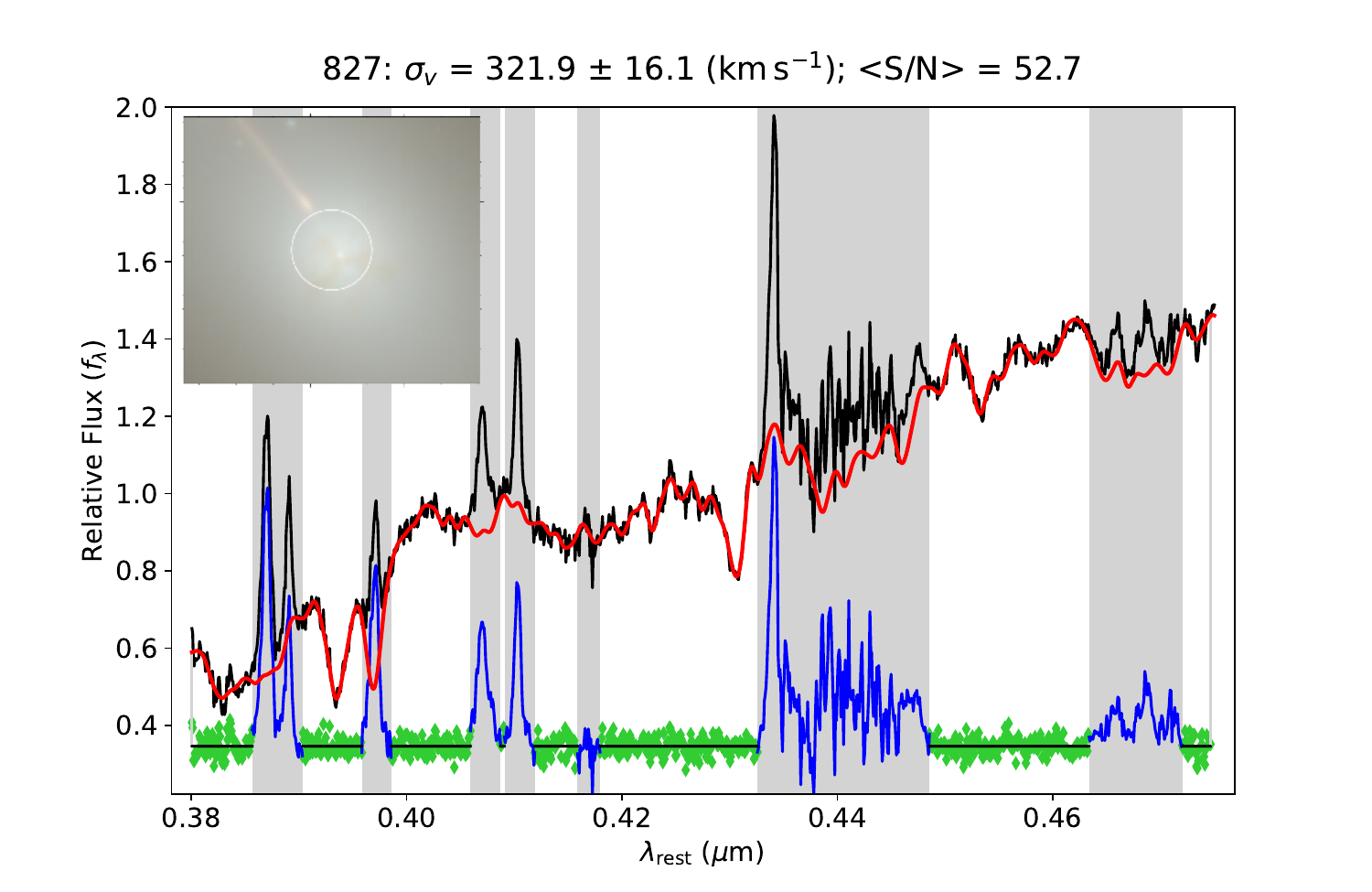}
  \includegraphics[width = 0.5\columnwidth]{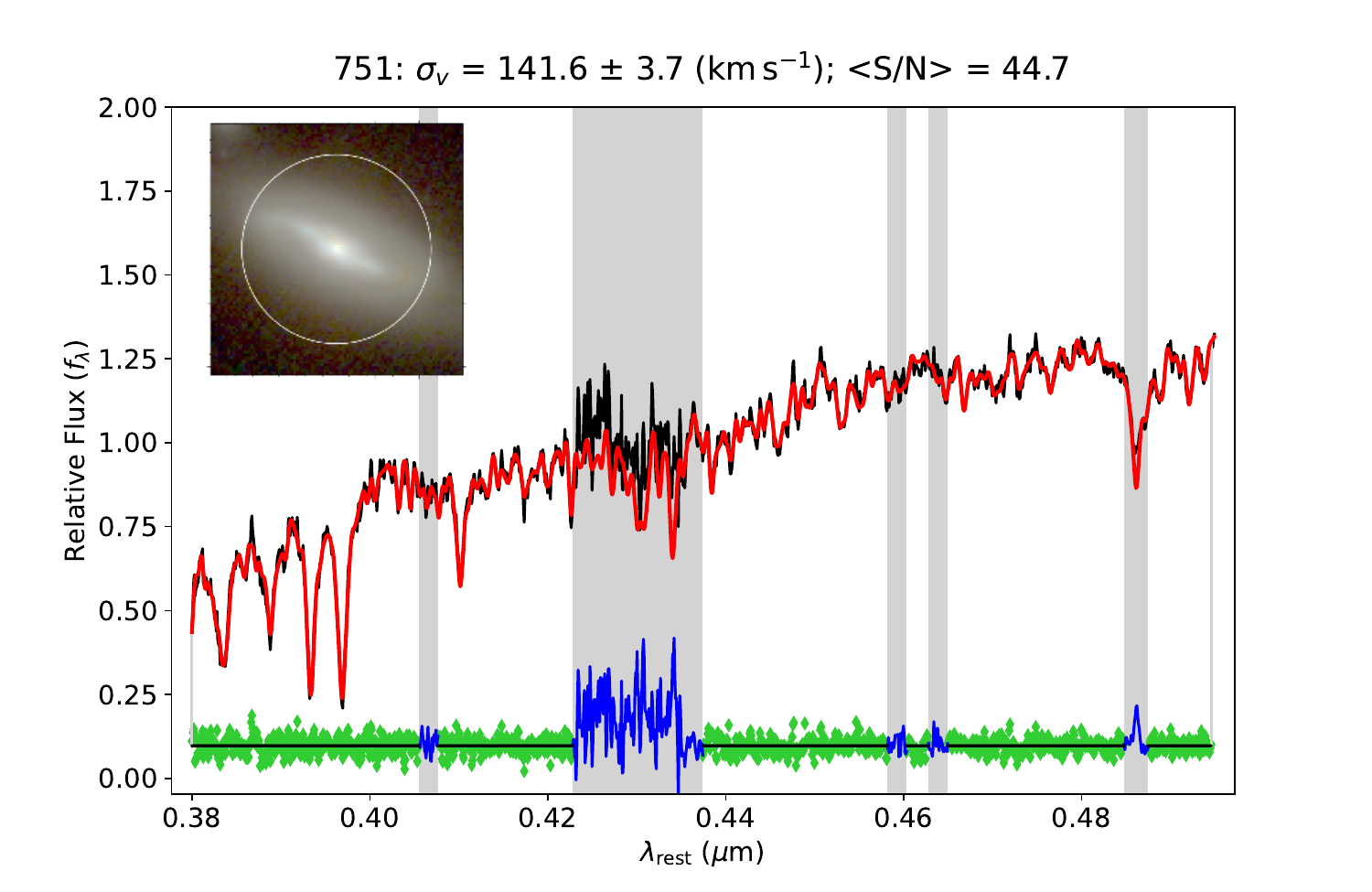}
  \includegraphics[width = 0.5\columnwidth]{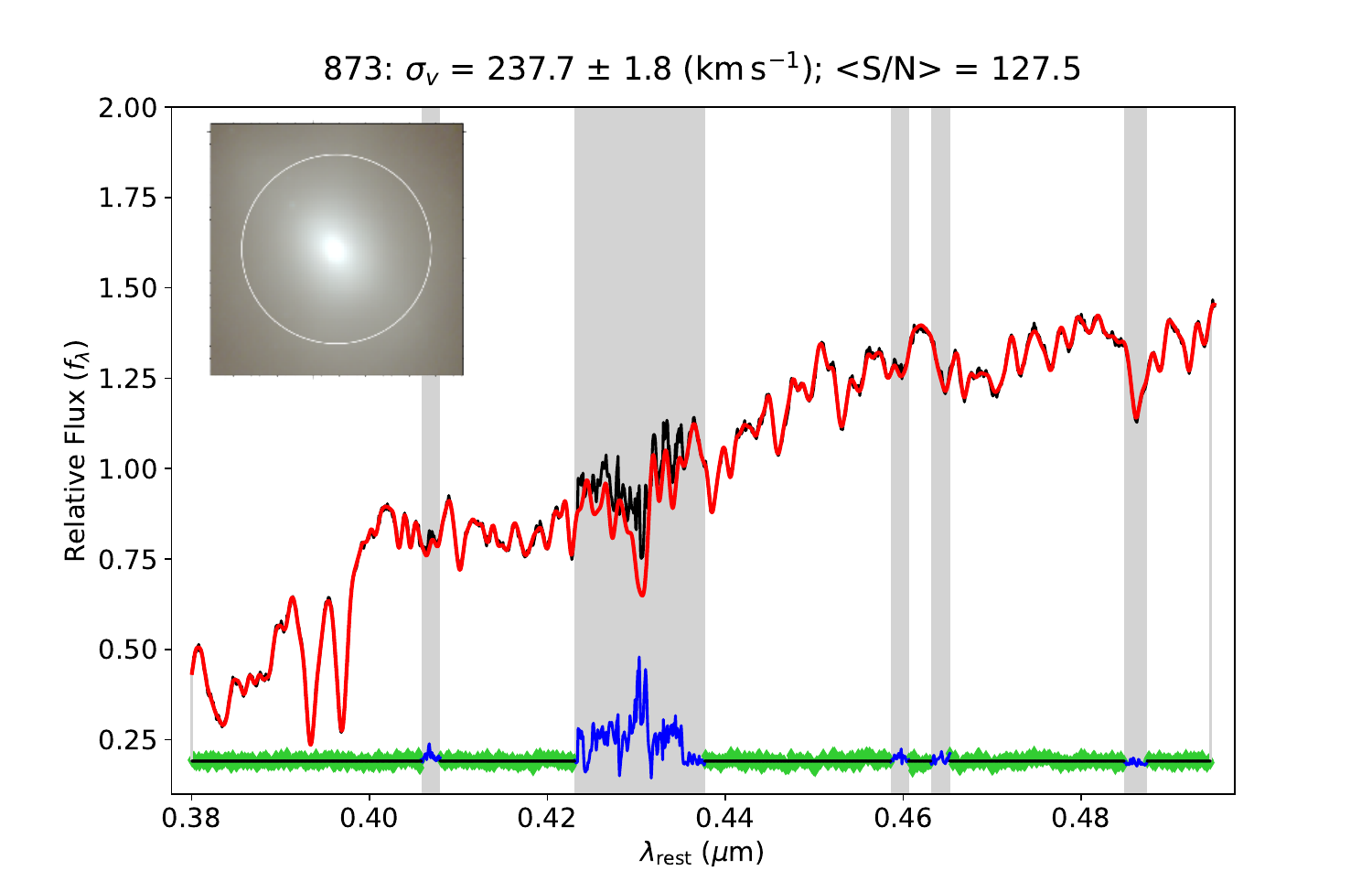}
  \caption*{Continued.}
  \label{fig:vdisp_spectra3}
\end{figure*}

\end{appendix}

\end{document}